\documentclass[AMA,STIX1COL]{WileyNJD-v2}
\usepackage{comment}
\articletype{Article Type}%

\received{DD Month YYYY}
\revised{DD Month YYYY}
\accepted{DD Month YYYY}

\begin{document}

\title{Uncertainty in Lung Cancer Stage for Outcome Estimation via Set-Valued Classification}

\author[1]{Savannah Bergquist*}

\author[2]{Gabriel Brooks}

\author[3]{Mary Beth Landrum}

\author[3]{Nancy Keating}

\author[4]{Sherri Rose}

\authormark{Bergquist \textsc{et al}}

\address[1]{\orgdiv{Haas School of Business}, \orgname{University of California, Berkeley}, \orgaddress{\state{CA}, \country{United States}}}

\address[2]{\orgdiv{The Dartmouth Institute for Health Policy and Clinical Practice}, \orgname{Geisel School of Medicine}, \orgaddress{\state{NH}, \country{United States}}}

\address[3]{\orgdiv{Department of Health Care Policy}, \orgname{Harvard Medical School}, \orgaddress{\state{MA}, \country{United States}}}

\address[4]{\orgdiv{Center for Health Policy and Center for Primary Care and Outcomes Research}, \orgname{Stanford University}, \orgaddress{\state{CA}, \country{United States}}}

\corres{*Savannah Bergquist. \email{bergquist@berkeley.edu}}

\presentaddress{2220 Piedmont Ave. \\Office F461 \\ Berkeley, CA 94720 }

\abstract[Summary]{Difficulty in identifying cancer stage in health care claims data has limited oncology quality of care and health outcomes research. We fit prediction algorithms for classifying lung cancer stage into three classes (stages I/II, stage III, and stage IV) using claims data, and then demonstrate a method for incorporating the classification uncertainty in outcomes estimation. Leveraging set-valued classification and split conformal inference, we show how a fixed algorithm developed in one cohort of data may be deployed in another, while rigorously accounting for uncertainty from the initial classification step. We demonstrate this process using SEER cancer registry data linked with Medicare claims data.}


%\jnlcitation{\cname{%
%\author{Bergquist S.}, 
%\author{G. Brooks}, 
%\author{M.B. Landrum}, 
%\author{N. Keating}, and 
%\author{S. Rose}} (\cyear{YEAR}), 
%\ctitle{Uncertainty in Lung Cancer Stage for Outcome Estimation via Set-Valued Classification}, \cjournal{Statistics in Medicine}, \cvol{YEAR;ISSUE:page1--pagelast}.}

\maketitle

\section{Introduction}\label{sec:intro}

Understanding generalizability across populations is important  for answering applied health questions\cite{Degtiar2021}, especially in the context of prediction algorithms.\cite{Steingrimsson2021} In many cases, prediction tools are developed for a specific cohort or population but have limited practical applicability because it is difficult to rigorously assess the uncertainty that comes from deploying a fixed prediction algorithm across settings. Population-level oncology research in the United States is an example of this phenomenon. Many new methods are developed to predict survival among individuals with cancer in Surveillance, Epidemiology, and End Results (SEER) cancer registry data, but often these tools lack the necessary information for replication in another study, even within other SEER cohorts.\cite{Hegselmann18} Similarly, there is great interest in predicting cancer stage using health care claims data because stage---an indicator of disease severity---is a key variable in understanding whether patients received appropriate treatments and for assessing health outcomes. The availability of valid algorithms to identify cancer stage using only claims data would allow for timely assessments of quality and outcomes as well as comparative effectiveness research in non-clinical trial settings. 

However, in applied classification settings, it can be a particular challenge to convey uncertainty. Often, a classification or risk prediction tool is developed with a focus on in-sample and validation sample performance measures, but little or no consideration is given to the use of the predicted values in subsequent estimation tasks.\cite{Hassett17,Ritzwoller18,Esposito19,Bergquist17,Brooks19b} Predicted values  are commonly treated as observed in downstream analyses. Depending on the specific applied context, this can lead to bias, deflated standard errors, and decreased interpretability.\cite{Wang20,Ogburn20}

To the best of our knowledge, there is no method that incorporates cancer stage classification uncertainty in survival estimation. The goal of this study is to develop a method that bridges between presenting raw probabilities---which may reflect our best sense of epistemic uncertainty albeit difficult to act upon in practice---and treating predicted class labels as ground truths. Using  SEER registry data linked to Medicare claims data, we develop prediction algorithms to classify patients as having stage I/II, stage III, or stage IV lung cancer. After evaluating predictive performance, we then examine survival outcomes stratified by these stage groups. Motivated by this real-world applied problem, we propose an approach that leverages set-valued classification, split conformal inference, and resampling to produce label sets that simplify underlying classification uncertainty and translate it into outcomes estimation. We describe this procedure and compare it to a naive classification and outcomes estimation method without  consideration of prediction uncertainty, and a naive method with uncertainty incorporated via resampling. In a simulation study, we  explore conditions under which the standard practice approach differs from the bootstrap-based alternatives, and show how uncertainty may vary across class labels. Both bootstrap methods are an improvement over standard practice in terms of taking steps to more comprehensively communicate uncertainty in applied settings. Our applied data example shows that in practice the simpler, naive bootstrap tool may perform as well as the more complex weighted labeling bootstrap procedure. 

\subsection{Related work on prediction and uncertainty}

The setting and related techniques we introduce are distinct from multiple imputation and the measurement error modeling literatures. Multiple imputation commonly seeks to generate values when a variable is missing for some but not all subjects in a data set. In cancer stage classification with claims data, and many other settings across medicine and health services research, the variable to be predicted is systematically missing for all observations in the  analysis sample. Similarly, covariate error measurement models often rely on some portion of observed data or external information pertaining to the variables measured with error, and require assumptions about the nature of the error (e.g., independent and additive). In our proposed approach we avoid the assumption that errors are independent of the predicted values and we do not require downstream users to observe any values for the variable to be predicted. 

Wang et al.\cite{Wang20} propose a method for correcting estimates, standard errors, and test statistics in settings where the research question of interest focuses on the relationship between a covariate and outcome, but the outcome must first be predicted. They term inference with predicted outcomes ``post-prediction inference.'' Their correction is based on the relationship between the observed and predicted outcome in a test data set. This post-prediction inference method is flexible but it is not developed for cases where the predicted variable is not the subsequent target outcome. 

In clinical risk prediction, a popular method for communicating uncertainty is generating an uncertainty score to accompany the predicted values.\cite{Meijerink20,Raghu19} The aim of this work is often to provide a measure of uncertainty to faciliate clinical decision-making, rather than using the predicted values and associated uncertainty measure in statistical analysis. However, these methods share a high-level goal with our proposed method and post-prediction inference: improve the effectiveness of prediction tools for applied problem solving. 

\subsection{Conformal inference and related methods}

The conformal inference framework was developed by Vovk et al.\cite{Vovk05} to provide an unbiased solution for obtaining a prediction interval around a new observation drawn from the same distribution as a set of training data. Note that a naive solution to creating a prediction interval  based on ranking the residuals from the training data is in practice unlikely to obtain target coverage, meaning the probability that a new test point lies in the prediction interval is less than 1-$\alpha$, where $\alpha$ is the specified error level. 

Split conformal inference separates the fitting and ranking steps so that the fitting only occurs as frequently as the data are split (i.e., with a single split, the prediction algorithm is fit once and the residuals are ranked once).  Papodopoulos\cite{Pap08}, Vovk\cite{Vovk12}, and Lei et al.\cite{Lei18} show that---assuming exchangeable, i.i.d. data and consistency of the regression estimator---split conformal inference yields prediction intervals with reliable conditional coverage. Health care data sets are typically subject to some gradual shifts in treatment practices and population demographics\cite{Nestor19}, and these types of drifts may or may not to threaten assumptions about the data distribution over limited time periods. However, investigator knowledge is critical for understanding events that can cause dramatic shifts and violate exchangeability (e.g., changes in medical coding regulations or the introduction of a blockbuster drug). Although split conformal inference can lead to wider prediction intervals than methods that use a greater number of splits, it is less computationally expensive and has been developed to work with classification outcomes. In this paper, we deploy split conformal inference to leverage these characteristics and extend it for purposes of generalizability. 

Although they are not directly applicable to our real-world data application of cancer stage classification and survival estimation because they do not produce label sets, we describe two further, related approaches in the literature that focus on producing prediction intervals for regression settings. Cross conformal inference\cite{Vovk15,Vovk18} is a natural extension of split conformal inference: instead of dividing the data in two halves, the data are divided into $V$ folds of equal size. Cross conformal inference guarantees 1-2$\alpha$ coverage when $V$ is small, but has no coverage guarantees when $V$ is large. Jackknife+ is a recently introduced method that uses leave-one-out predictions for the prediction interval (unlike the conventional jackknife, which uses the full training data).\cite{Barber2019} Barber et al.\cite{Barber2019} also propose a method called CV+, which yields larger intervals than jackknife+ but requires less computation. We did not pursue adapting these approaches in this study because we are focused on an applied classification setting. 

\section{Methods}\label{sec:methods}

\begin{figure}[htbp]
	\centering
	\includegraphics[scale=.6]{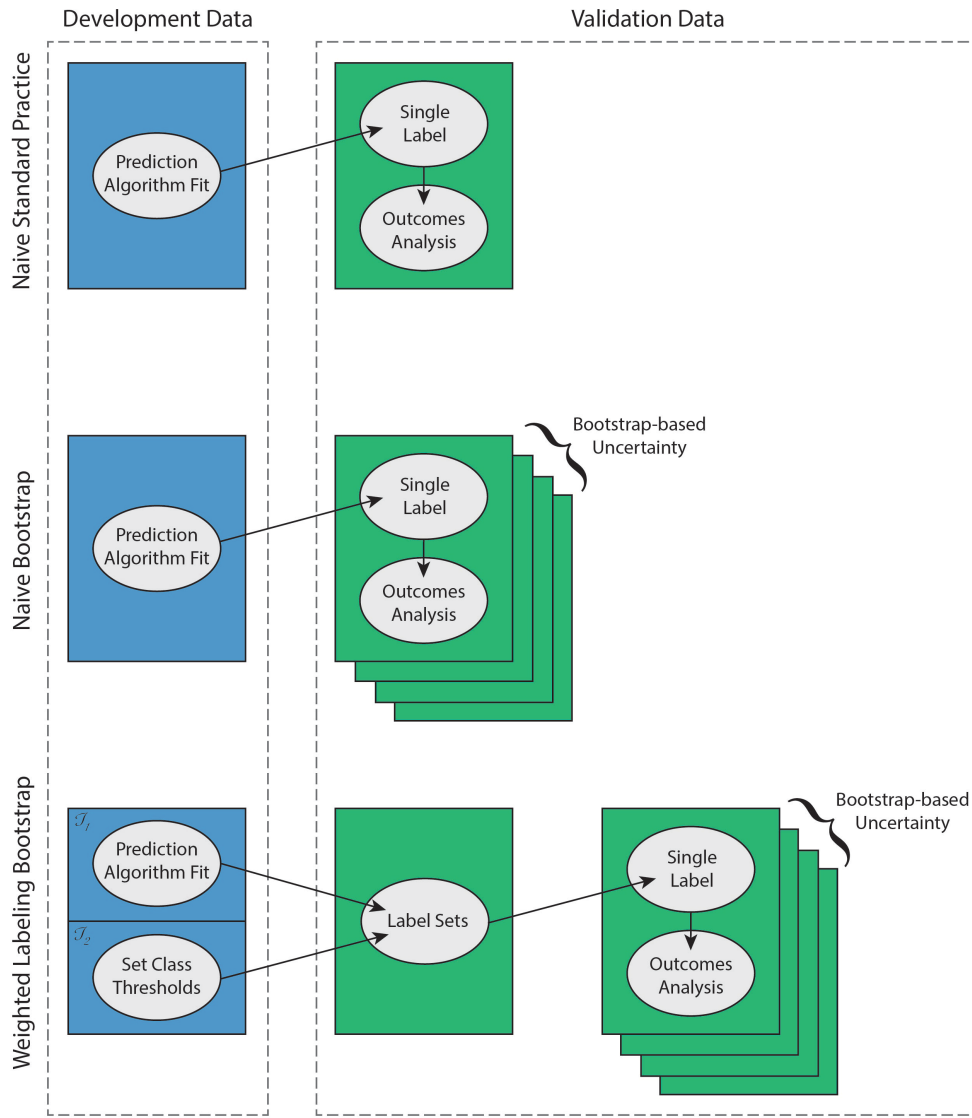}
	\caption{Conceptual overview of methods.}
	\label{fig:overview}
\end{figure}

We will compare three methods, which are summarized conceptually in Figure~\ref{fig:overview}. In the naive standard practice approach, the prediction algorithm is fit in the development data and applied to a validation data set, yielding a single label per observation that is used in subsequent outcomes analysis. In our data setting, the naive approach is to apply a fitted classification algorithm in the validation data, assign lung cancer stage based on the greatest predicted class probability, and then use this stage label to estimate survival or other outcomes of interest. Under this naive approach, no special consideration is given to the uncertainty from predicting class labels. In the naive bootstrap method, uncertainty is incorporated by bootstrapping the validation data. This resampling allows for bootstrap-based confidence intervals around classification performance measures and outcomes estimation. Finally, the weighted labeling bootstrap procedure divides the development data into two halves to leverage split conformal inference (discussed in more detail below). The fitted algorithm and label thresholds are then applied to the validation data to assign a set of plausible labels for each individual observation. The validation data are resampled and a single label is selected from each label set and used in outcomes estimation. Thus the weighted labeling approach captures classification and outcomes estimation uncertainty via set-valued classification (the label sets) and resampling. 

\subsection{LABEL classifier and weighted bootstrap}

Our data structure is defined as $O=(X,Y)$, where $X$ is a feature vector and $Y=\{1,...,K\}$ is the mutually exclusive label space. In the multiclass setting, we seek to apply a label $y \in Y$ to each observation based on $x \in X$. We set $K\!=3$; in our applied data analysis, lung cancer stage I/II, stage III, and stage IV map to labels 1-3, respectively. Although each observation has a single true label, our goal is to assign a \emph{set} of plausible labels in order to provide more information about the uncertainty of the classification process. To do so, we employ a set-valued classifier proposed by Sadinle et al.\cite{Sadinle19}, the least ambiguous with bounded error levels (LABEL) classifier. In combination with split conformal inference, the LABEL classifier yields distribution-free, finite sample coverage levels with minimal label ambiguity when the assumption of exchangeability holds. To estimate each $y$, we require an initial estimate of the conditional probability function, $P(y|x)$, and class-specific thresholds for converting probabilities to labels, $\{t_y\}^K_{y=1}$. Any conventional estimator of $P(y|x)$ can be plugged in to the LABEL classifier; in our simulation study we consider a multinomial logistic regression and in our data analysis we include a range of estimators.

We desire equal coverage across all three lung cancer stage groupings, and therefore pre-specify individual class error levels of $\{\alpha_y\}^K_{y=1}= 0.10$. As described in Sadinle et al.\cite{Sadinle19}, in the optimal procedure, the set-valued classifier, $\mathbf{H}(X)$, is a subset of the label space and can be represented by a collection of sets mapping the feature inputs to labels: $C_y =\{  x \in X : y \in \mathbf{H}(X)\}$. 
 The estimated label sets for each class label, $\widehat{C}_y = \{x : \widehat{P}(y|x) \geq \widehat{t}_y \}$, are based on the estimated conditional probabilities, $\widehat{P}(y|x)$, and estimated thresholds, $\widehat{t}_y$. 

 The coverage for class $y$ is based on the collection of estimated label sets, $\widehat{C}_y$, that contain the true class label:
\begin{displaymath}
	\widehat{\text{coverage}}_y(\widehat{t}_y) = \frac{\sum_{i=1}^n I(X_i \in \widehat{C}_y)I(Y_i=y)}{\sum_{i=1}^n I(Y_i=y)}, 
\end{displaymath} 
\noindent where $i$ indexes $n$ independent observations from $O$. The threshold $\widehat{t}_y$ is estimated: 
\begin{eqnarray}\nonumber
	\widehat{t}_y= \max_{i: Y_i=y}[\widehat{P}(Y_i|X_i): \widehat{\text{coverage}}\{\widehat{P}(Y_i|X_i)\} \geq 1-\alpha_y]
\end{eqnarray}

In an applied, finite sample setting with split conformal inference, the data are split in two halves, indexed by $\mathcal{I}_1$ and $\mathcal{I}_2$. For our application, the data are first divided into a development and validation cohort, and the development cohort is split into $\mathcal{I}_1$ and $\mathcal{I}_2$. The $\mathcal{I}_1$ subset is used to estimate $\widehat{P}(y|x)$. We partition $\mathcal{I}_2$ into $K$ groups corresponding to each label class: $\mathcal{I}_{2,y} = \{i \in \mathcal{I}_2 : Y_i = y\}$. Within the given partition $\mathcal{I}_{2,y}$, thresholds $\widehat{t}_y$ are estimated according to pre-specified error levels $\alpha_y$:
\begin{eqnarray}\nonumber
	\widehat{t}_y =
	\min_{i \in \mathcal{I}_{2,y}} \Biggl\{ \widehat{P}(Y_i|X_i): \sum_{j \in \mathcal{I}_{2,y}} I\{\widehat{P}(Y_j|X_j) \leq 
	\widehat{P}(Y_i|X_i)\} > (|\mathcal{I}_{2,y}| + 1)\alpha_y-1  \Biggr\}, \phantom{lor} 
\end{eqnarray}

where $j$ indexes over $\mathcal{I}_2$ for comparison of each predicted probability to the ranked predicted probabilities. We then obtain the label sets $\widehat{C}_y = \{x : \widehat{P}(y|x)\geq \widehat{t}_y\}$ so that $\{\widehat{C}_y\}^K_{y=1}$ is the split conformal set-valued classifier based on the plug-in estimator $\widehat{P}(y|x)$. Instances where more than one label is assigned to an observation are called ambiguous sets, and label sets with no assigned labels are null sets. With $\alpha_y$ set equally across classes, a higher threshold for a given class $y$ indicates a greater number of higher predicted probabilities generated by $\widehat{P}(y|x)$. We apply the classifier---based on the plug-in estimator fit in $\mathcal{I}_1$ and the thresholds obtained in $\mathcal{I}_2$---to the observations in the validation cohort. Separating the creation of the classifier from the validation cohort allows the classifier to be used in data settings where the true class is unobserved. 

Our end goal is to use the class labels in outcome estimation. Thus, we force a single label per observation in a bootstrap procedure: for all $n_{\text{boot}}$ resamples of the validation data, a single label, $\widehat{y}_{boot}$, is randomly selected with equal probability from all classes $y$ where $\widehat{P}(y|x) \geq \widehat{t}_y$. For null sets, all class labels are considered with equal probability (e.g., in settings where the classification algorithm has high average accuracy and there is little uncertainty, we may expect higher classification thresholds and thus null sets for difficult-to-classify observations). After a single label is randomly selected among those belonging to the label sets $\widehat{C}_y$, outcome estimation can be performed based on $\widehat{y}_{boot}$. We refer to this approach as ``weighted'' labeling with bootstrap, or weighted bootstrap, because the probability of label assignment for a resampled observation is weighted by the label set obtained prior to the resampling procedure. For example, in the case of $K=3$, the set of possible label assignment weights is $\{0,0.33,0.5,1\}$. 

The bootstrap procedure allows the class labels assigned to a single observation to vary, thus incorporating classification uncertainty in both classification evaluation and outcomes estimation. Additionally, the resampling permits us to calculate confidence intervals for a variety of metrics of interest, including label coverage, counts of ambiguous label sets, conventional classification evaluation measures, and outcome measures. Because there is no sample mean for metrics that depend on a single class per observation (for example, classification accuracy), we report bootstrap percentile-based averages and confidence intervals.

\begin{boxtext}
\noindent \underline{\textbf{Algorithm: Weighted labeling with bootstrap.}\hspace{.612\linewidth}}
\begin{enumerate}[1.]
	\item For development data:
	\begin{enumerate}[a.]
		\item Split development data into equal halves: $\mathcal{I}_{1}$ and $\mathcal{I}_{2}$
		\item Fit conditional probability estimator $P(y|x)$ on $\mathcal{I}_{1}$ 
		\item Apply $\widehat{P}(y|x)$ to $\mathcal{I}_{2}$ to obtain thresholds $\widehat{t}_y$
	\end{enumerate}
	\item For validation data:
	\begin{enumerate}[a.]
		\item Apply $\widehat{P}(y|x)$ and $\widehat{t}_y$ to obtain label sets $\widehat{C}_y$ and calculate coverage
		\item Draw $n_{\text{boot}}$ resamples:
		\begin{enumerate}[i.]
			\item Select a single label $\widehat{y}_{boot}$ from each resampled observation's label set
			\item Calculate classification performance measures based on $\widehat{y}_{boot}$
			\item Estimate outcome of interest based on $\widehat{y}_{boot}$
		\end{enumerate}
		\item Report statistics of interest based on validation sample and validation bootstrap resamples
	\end{enumerate}
\end{enumerate}
\vspace{-10pt}
\end{boxtext}

\subsection{Naive comparators: Most probable class, with and without resampling}

As comparators, we implement a naive classification strategy where a single class label is assigned to each observation based on the most probable predicted class. Both naive comparators implement this classification approach. We use the same conditional probability estimator as in the weighted labeling approach, $P(y|x)$, but fit it on the entire development cohort. The fitted classification algorithm is then applied to the validation cohort, and class labels are assigned based on the class with the highest predicted probability. These class labels are treated as known in subsequent outcomes estimation exercises. We use this approach, when applied in a single iteration of an analysis sample, as our standard practice comparator. A step beyond implementing this naive standard practice strategy is to incorporate bootstrap resampling. The classification algorithm fitting and prediction steps remain unchanged, but a bootstrap procedure is introduced for the classification performance evaluation and outcome estimation: the validation cohort is resampled and bootstrap-based confidence intervals may be calculated for classification and outcome measures. We include the naive bootstrap approach as a more nuanced method than the naive standard practice, but a less complex alternative to the weighted labeling bootstrap procedure introduced above. Depending on the real-world data context, the simpler, naive bootstrap may be preferable to the weighted labeling method.

\subsection{Classification performance measures}

For a given $\widehat{P}(y|x)$, we assess classification performance according the measures in 
Table~\ref{tab:measures}. To calculate these measures, we tabulate the per-class proportions of  true positives, true negatives, false positives, and false negatives. We then present what we call the macro-averaged versions of each measure by averaging across the classes. Because the split conformal inference-based LABEL classifier does not provide any coverage guarantees in this resampling setting, we empirically verify coverage and examine estimated thresholds and ambiguous labels for each class. In our data application, we also plot the proportion of observations in a given stage by ordered predicted probability to assess calibration. 

\section{Simulation Study}\label{sec:simstudy}

In this section, we describe a simulation study to compare the methods for classification and outcome estimation outlined above. The goal is to highlight settings where the relative performance of a naive, single iteration classification and outcome estimation approach (``standard practice'') varies from a similar naive approach that incorporates a bootstrap procedure and from our proposed novel approach of weighted labeling with bootstrap, which leverages set-valued classification. Although we construct our simulation data to be similar to our real-world lung cancer data setting, the simulation study allows us to explore methodological performance under differing data scenarios. To reflect this, throughout the simulation study we refer to the classification outcome variable groups as ``classes'' rather than stage.

Our simulation sample is $N=2000$ observations and we generate patient-level covariates $\mathbf{X}_i$ for each patient $i$, where $i={1,...,N}$. To roughly approximate the sociodemographic and clinical characteristics of the lung cancer study, we simulate six continuous, seven binary, and two count variables; Table~\ref{tab:appsim} provides additional details on covariate distributions. The class label outcome $Y=\{1,...,K\},$ where $K=3$, is generated from a multinomial distribution, $(Y_1,Y_2,Y_3) \sim \text{Multinom}(N,(p_1,p_2,p_3))$ where the class label $Y_i$ is assigned based on the greatest class probability. Class balance is approximately 37\% class 1, 49\% class 2, and 13\% class 3. See Appendix \ref{app1} for additional details on the construction of the probabilities underlying $Y$.  

Survival outcomes are central to many strands of oncology research. To examine how our method compares to the naive approaches across simulation scenarios, we implement a common form of survival estimation, stratified Kaplan-Meier analysis. We select Kaplan-Meier estimation for simplicity and ease of exposition, but many alternative methods for survival outcome estimation could be used in practice, including standard parametric survival regression models, the Cox proportional hazards regression\cite{Cox1972}, random survival forests\cite{Ishwaran2008}, and combinations of parametric, semi-parametric, and nonparametric approaches in a stacked survival model.\cite{Hothorn2006, Wey2015, Golmakani2020}

Using a Weibull distribution, we simulate right censored event times: $T_i = (-\log(U_i) b \exp(-Y_{i}\beta_k))^{\frac{1}{a}}$, where $T_i$ is simulated survival time influenced by time-independent covariate $Y_{i}$ with effect parameter $\beta_k$, $U_i \sim \text{Uniform}(0,1)$, $a$ is the Weibull shape parameter, and $b$ the scale parameter. We set $a=1$, $b=90$, and $Y_i$ is the class label simulated for the classification exercise. We set the length of observation time to 365 days and generate a variable to capture observed censored event times and an indicator if the event occurred or was censored. After simulating all variables, the data are stratified on the class label $Y_i$ and divided into equally sized development and validation cohorts, each of size 1000. 
For the weighted labeling with bootstrap method, the development data are again stratified on $Y_i$ and further divided into two halves of 500 observations each. % Data are simulated using R 3.6.1 using a Mersenne-Twister random number generator with an input seed ``33.''

We implement three scenarios: 1) accurate and certain classification, 2) accurate and uncertain classification, and 3) inaccurate and uncertain classification. Table~\ref{tab:appsim} compares the covariates used to fit the multinomial logistic regression  for label classification in each scenario; scenarios differ only by which covariates are used to predict class label probabilities; all other simulation parameters are the same. We characterize the scenarios based on how accurately the classification regression predicts class labels with the given covariates, and based on the empirical distributions of the predicted probabilities (the uncertain scenarios have a higher proportion of predicted probabilities near 0.5). After fitting the conditional probability estimation algorithms and applying labels in the validation data, Kaplan-Meier survival estimation is performed in the validation data, stratified by observed and predicted class labels. We draw 500 bootstrap resamples for the bootstrap-based methods and we perform 1000 simulation repetitions for each scenario.

Across the simulation scenarios, we compare how each of the three methods performs in terms of classification  and survival analysis. For classification performance we examine accuracy, sensitivity, specificity, and positive predictive value (PPV), as well as the underlying counts of true positive, true negatives, false positives, and false negatives. We present mean measures and bootstrap-based 95\% confidence intervals where applicable. For all methods, we report coverage and 95\% confidence intervals for each class. We also report the class-specific thresholds and ambiguity for the weighted labeling classifier (recall the naive methods do not incorporate threshold-based label assignment and only assign a single label per observation so there is no label ambiguity).

Our primary survival analysis performance measure is bias: how do the median and 90 and 365-day survival probabilities based on predicted class compare to those based on observed class? We examine the average bias across simulations, report 95\% confidence intervals and present standard deviation of the bias. For the single iteration survival estimates (observed class and standard practice predicted class), confidence intervals are calculated as survival $\pm 0.95 \times \text{sd}(\text{survival})$. Percentile-based confidence intervals are calculated and we report means of the bootstrap samples for the bootstrap-based methods.

\subsection{Simulation results}

Class-specific coverage results across simulation scenarios are presented in Table~\ref{tab:appsimcov}. The weighted bootstrap method yields 90\% coverage for all three classes across settings. Note that although we do not fill in null sets until the bootstrap procedure, because the number of null label sets is low (described below), our classifier is still able to obtain the target coverage level. The naive classification method of assigning class labels based on the highest probability produces coverage ranging from $<$1-94\% across all three classes, and coverage declines within each class from scenario 1 (accurate and certain class prediction) to scenario 2 (accurate and uncertain) to scenario 3 (inaccurate and uncertain). The naive classification methods do not target coverage, while class-specific target coverage is an explicit input of the LABEL classifier in the weighted bootstrap method. The two naive methods are identical at this stage because coverage is calculated in sample prior to the bootstrap procedure.

\subsubsection{Classification performance}
The classifier thresholds in the weighted labeling with bootstrap method tend to decrease within class as uncertainty and inaccuracy are introduced across simulation scenarios (Table\ref{tab:appsimthlds}). Recall the thresholds are class-specific cutoff probabilities for determining if an observation receives a given class label. Class 2 is the most prevalent class (approximately 49\% of the sample), class 1 is approximately 37\% of the sample, and class 3 comprises approximately 13\%. In scenario 1, the most accurate and certain setting, class 1 has the highest threshold, illustrating that the most prevalent class will not necessarily generate the highest label threshold.

\begin{figure}[htbp]
	\centering
	\includegraphics[scale=.6]{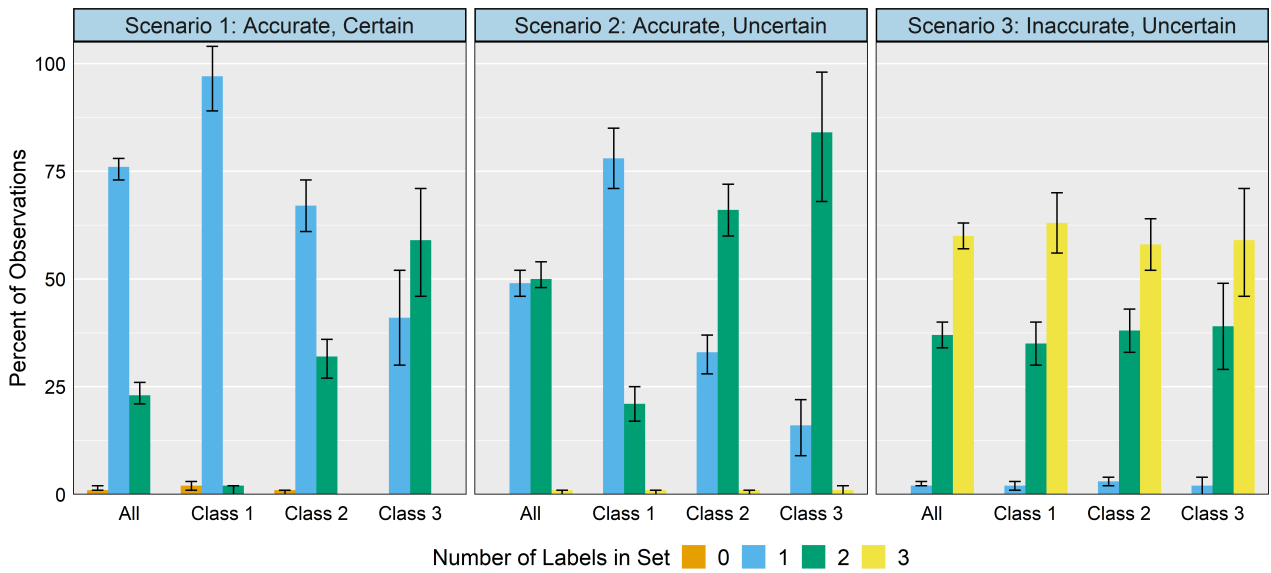}
	\caption{Simulation study label ambiguity: Share of sample by number of assigned labels in label set.}
	\label{fig:simamb}
\end{figure}

Label set ambiguity provides a more complete picture as to how the classifier operates (Figure~\ref{fig:simamb}). As the accuracy and certainty declines across simulation scenarios, we see the number of null and single label sets decrease and the number of double and triple label sets increase. Examining ambiguous label sets by true class, we see where the distribution of label sets differs by class. For example, in scenario 1 (accurate and certain), about 75\% of all observations receive a single label set. But for class 1, over 95\% of observations have a single label set, while in class 2 only 67\% have a single label and in class 3 the proportion drops to 41\%. Under scenario 2 (accurate and uncertain), 49\% of all observations receive a single label set, but again this differs by true class: 78\% of class 1 observations have a single label, 34\% of class 2, and 21\% of class 3. Under the inaccurate and uncertain scenario 3, the distribution becomes more even, with 2\% of class 1 observations having a single label, 3\% of class 2, and 2\% of class 3. Label set ambiguity can be helpful as a diagnostic tool, particularly in the multiclass setting where uncertainty levels may vary across classes. 

\begin{figure}[htbp]
	\centering
	\includegraphics[scale=.7]{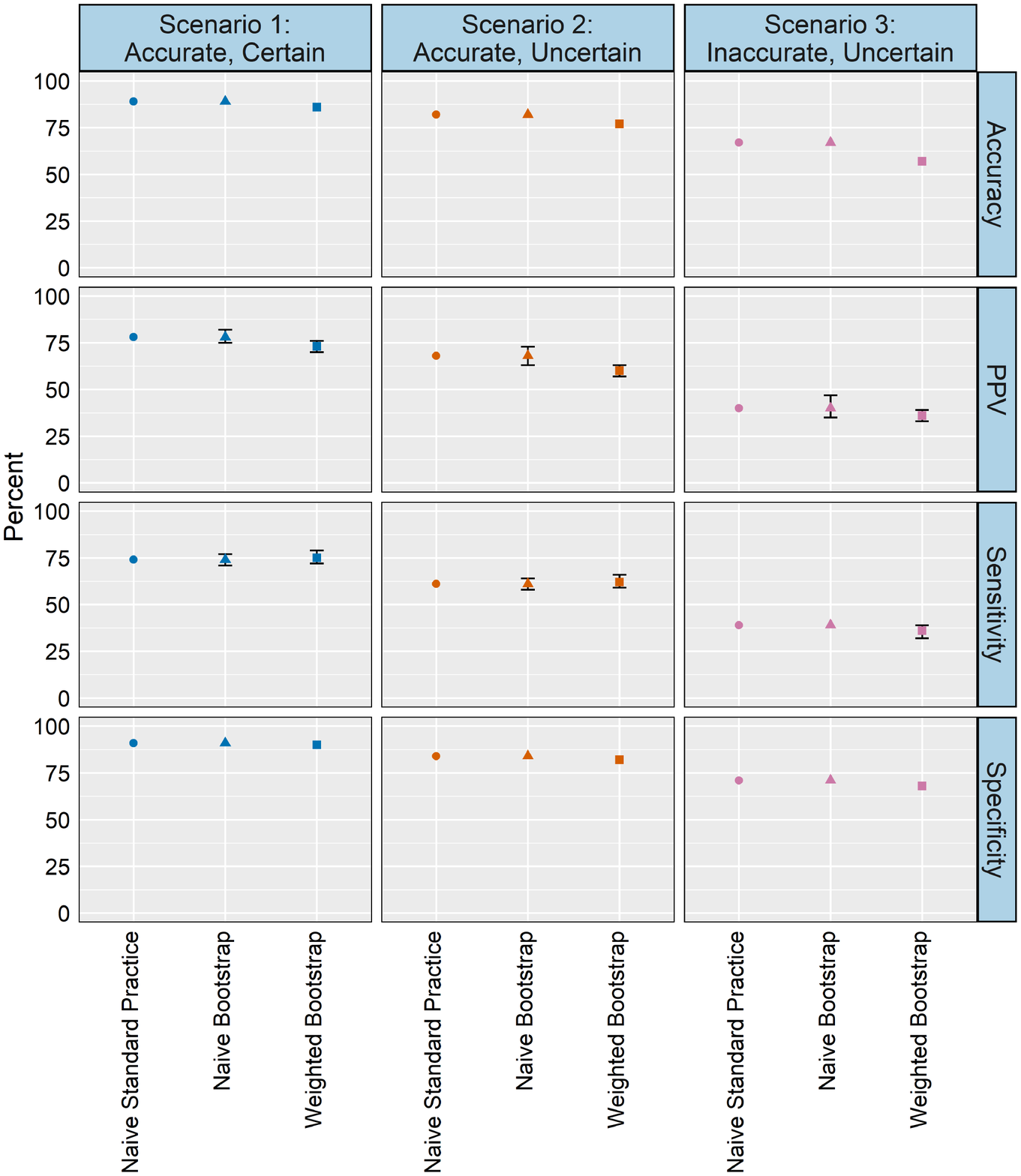}
	\caption{Simulation study classification discrimination: Average accuracy, sensitivity, specificity, and PPV.}
	\caption*{(\emph{For visual clarity, 95\% confidence intervals less than 0.05 are not displayed.})}
	\label{fig:appsimcm}
\end{figure}

Figure~\ref{fig:appsimcm} reports average classification accuracy, sensitivity, specificity, and PPV across simulation scenarios. Overall, the weighted labeling approach yields similar performance across metrics compared to the naive methods. Small differences can be seen for some metrics.  For example, in scenario 2, weighted labeling has an accuracy of 77\% while the naive methods have an accuracy of 82\%. However, weighted labeling has higher sensitivity (67\% vs. 61\%). Such differences may or may not be meaningful depending on the application. Table~\ref{tab:appsimacc} reports class-specific classification accuracy. Results for class 1 are similar across the methods, but the weighted labeling performs worse compared to the naive methods in classes 2 and 3 by 4--7 percentage points. 

Tables~\ref{tab:appsimsens}-\ref{tab:appsimppv} contain class-specific classification sensitivity, specificity, and PPV. The small differences between the naive methods and the weighted bootstrap are much more pronounced at the class-level. For example, the weighted bootstrap produces lower sensitivity for class 2 (by 12 percentage points in scenario 1, and about 25 percentage points in scenarios 2 and 3), but performs substantially better for class 3. In scenario 1, the weighted bootstrap sensitivity estimates are a 20 percentage point improvement over the naive methods, and an approximately 30 percentage point improvement in scenarios 2 and 3 (in scenario 3, the naive methods yield 0\% sensitivity). Tables~\ref{tab:appsimtp}-\ref{tab:appsimfn} present true positive, true negative, false positive, and false negative counts by class across simulation scenarios. The conditional probability estimates for class 3 are generally low compared to classes 1 and 2, leading to few class 3 labels under the naive approaches. Across all settings, the weighted bootstrap method generates fewer true positives than the naive methods for classes 1 and 2, but yields a substantially larger number for class 3. The practical import of these class-based differences will vary by applied context, especially considering our primary outcome of interest here is survival.

\subsubsection{Survival analysis}
\begin{figure}[htbp]
	\centering
	\includegraphics[scale=.6]{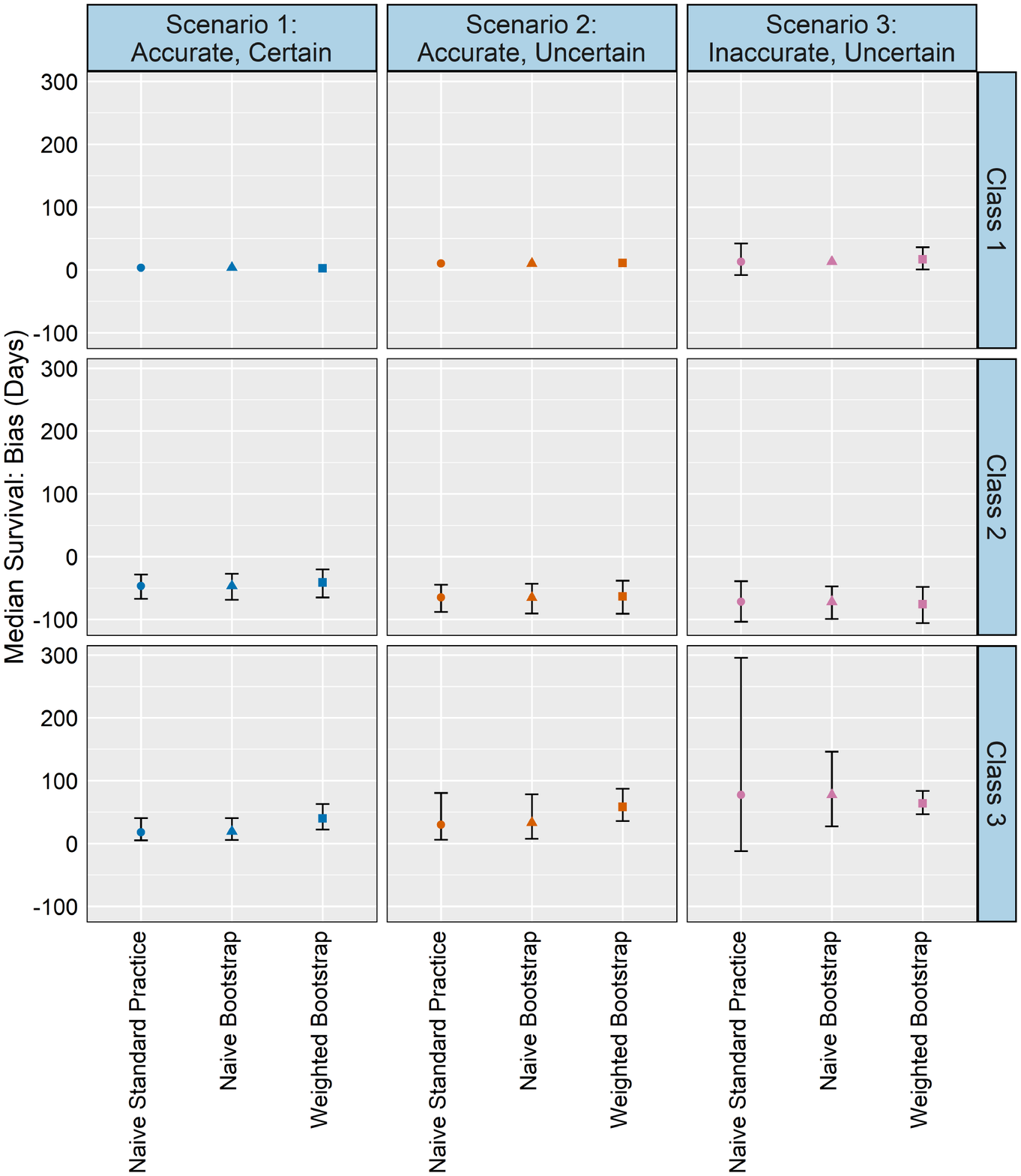}
	\caption{Simulation study median survival days bias.}
	\caption*{(\emph{For visual clarity, 95\% confidence intervals less than 31 days are not displayed.})} % CI < 31 days
	\label{fig:simmedbias}
\end{figure}

\begin{figure}[htbp]
	\centering
	\includegraphics[scale=.6]{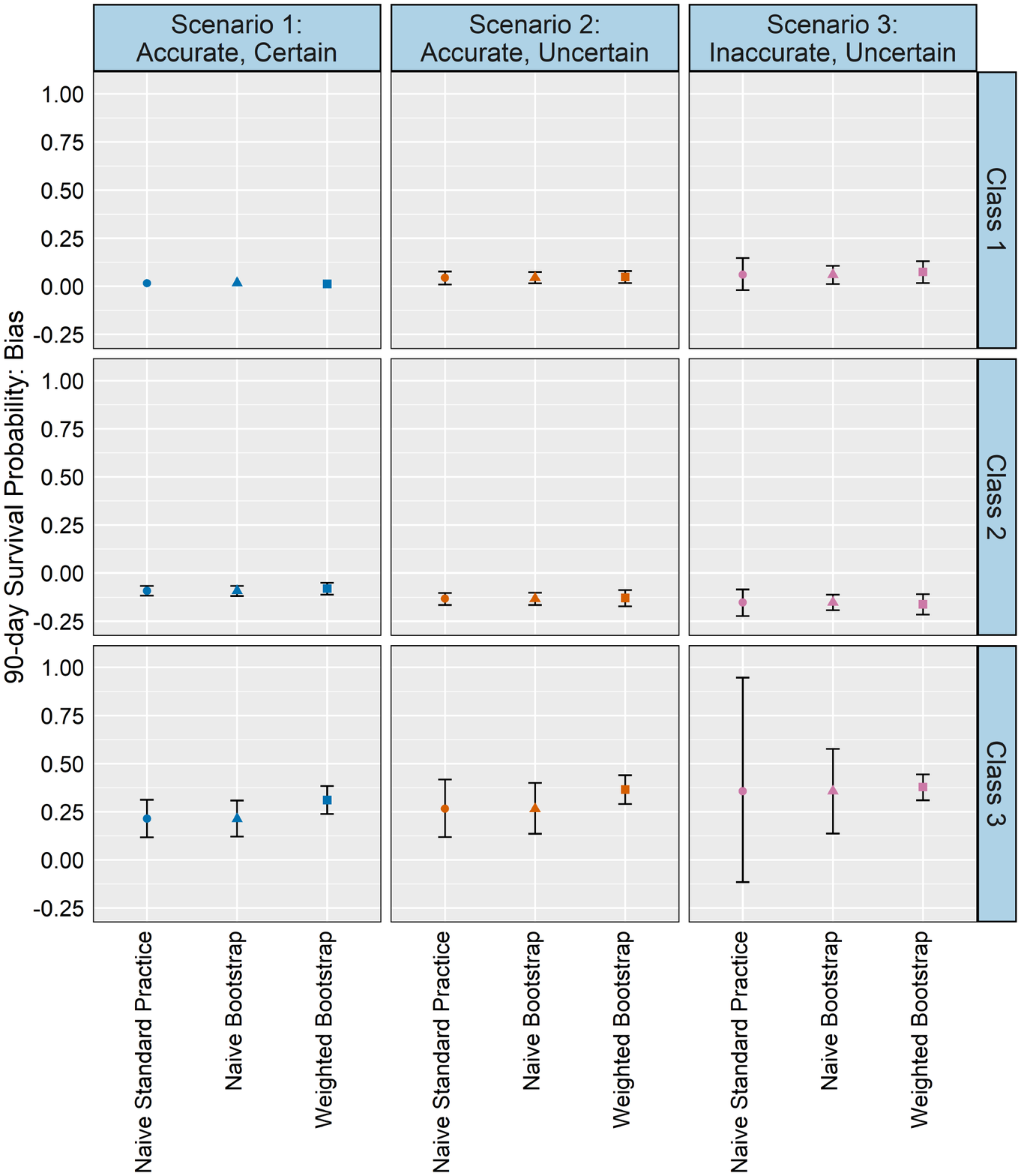}
	\caption{Simulation study 90-day survival probability bias.}
	\caption*{(\emph{For visual clarity, 95\% confidence intervals less than 0.05 are not displayed.})}
	\label{fig:appsim90bias}
\end{figure}

\begin{figure}[htbp]
	\centering
	\includegraphics[scale=.6]{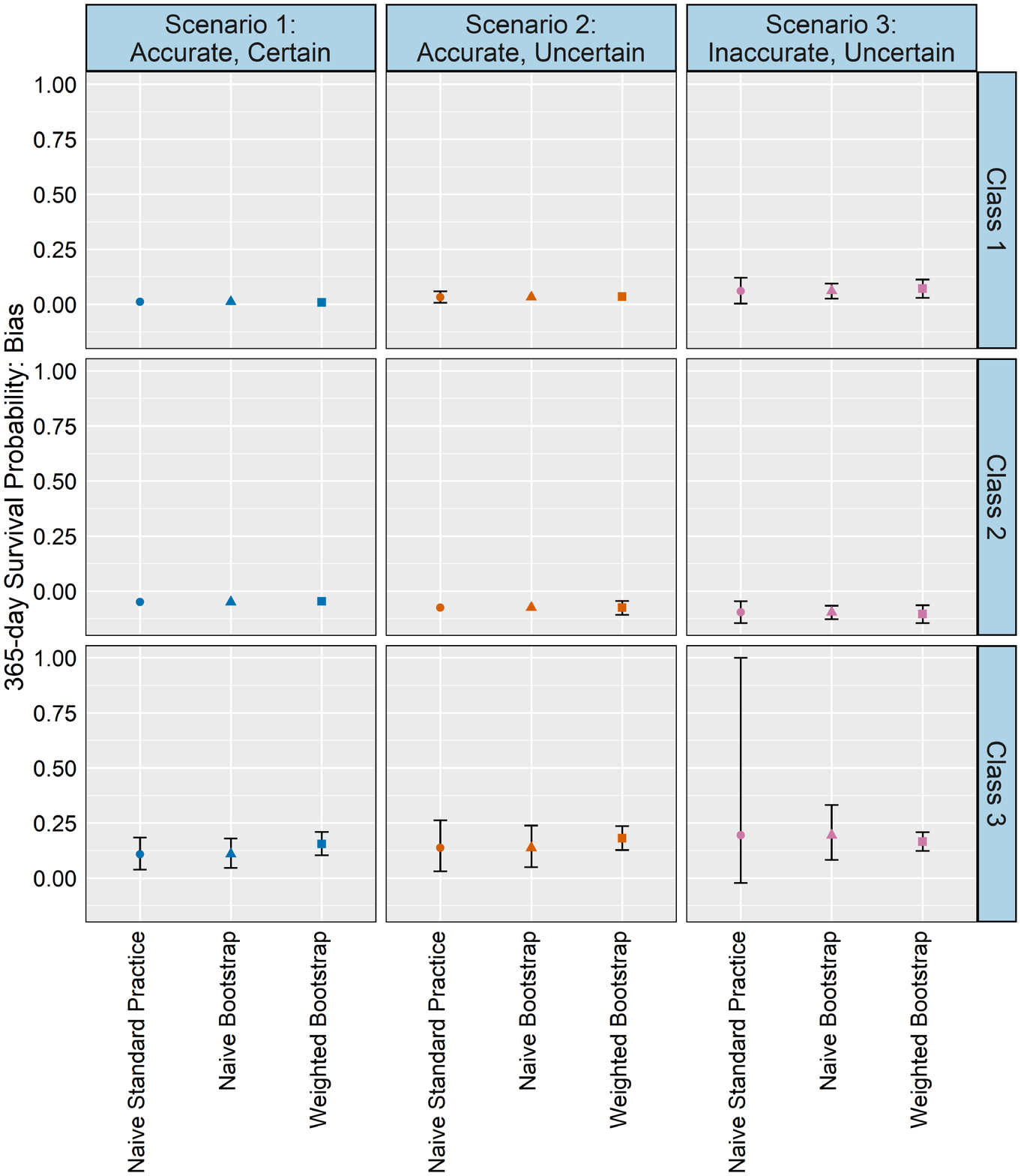}
	\caption{Simulation study 365-day survival probability bias.}
	\caption*{(\emph{For visual clarity, 95\% confidence intervals less than 0.05 are not displayed.})}
	\label{fig:appsim365bias}
\end{figure}

Bias for median survival is shown in Figure~\ref{fig:simmedbias}. Overall, the weighted bootstrap produces smaller bias about 40\% of the time compared to the naive methods. The weighted bootstrap and naive methods have similarly sized confidence intervals across scenarios for classes 1 and 2. For class 3, the weighted bootstrap has a much smaller confidence interval than the naive methods, particularly so in scenario 3. The results for 90 and 365-day survival (Figures\ref{fig:appsim90bias}, \ref{fig:appsim365bias}) run counter to what one might expect: rather than yielding smaller standard errors or less biased average survival probabilities than the naive or weighted bootstrap methods, standard practice produces nearly identical results, with the noted exception of class 3 in the inaccurate and uncertain scenario.

\section{Data Analysis}\label{sec:data}

We use SEER cancer registry data linked with Medicare claims data as our real-world data application.\cite{Enewold20} The SEER data provide information on all cancers diagnosed among individuals living in areas covered by SEER registries, including cancer stage.\cite{SEER2019} SEER data are abstracted from medical records and contain validated staging information at the time of diagnosis, and thus are the source of our ``gold standard'' cancer stage labels: stage I/II, stage III, and stage IV. We combine stages I and II into a single label to accommodate sample size constraints and similarity of clinical outcomes.\cite{SEER2019} Fee-for-service Medicare claims data contain detailed information on treatments received as well as health care visits and comorbidities. Medicare enrollment data provide information on patient age, race/ethnicity, vital status, and information on zip-code level measures of socioeconomic status.

Our study cohort includes individuals aged 65 and older who were enrolled in fee-for-service Medicare and were diagnosed with stage I-IV lung cancer between 2010 and 2013 who received chemotherapy within 6 months of diagnosis. We divide the data into two cohorts based on timing of diagnosis: a development cohort (2010--2011 diagnoses) and a validation cohort (2012--2013 diagnoses). In practice a fixed classification algorithm is likely to be applied to individuals diagnosed in time periods successive to the training sample, thus we aim to approximate this likely real-world scenario.  Table~\ref{tab:cohorts} shows the similarity between the two cohorts in terms of basic summary statistics; the similarity of our development and validation samples can be considered a ``best case'' scenario in terms of generalizability and prediction. 

\begin{center}
\begin{table*}[htbp]
\centering
	\caption{Cohort summary statistics.\label{tab:cohorts}}%
	{\begin{tabular}{ l r r } 
			\toprule
			\multirow{2}{*}{\bf{Characteristic}} & \bf{Development Cohort} & \bf{Validation Cohort}\phantom{jj}  \\
			& \bf{2010-2011 Diagnosis}\phantom{l}  & \bf{2012-2013 Diagnosis} \\ 
			\midrule
			N & 14,760 & 14,620 \\
			Age (mean) & 72.1 & 71.9 \\
			Documented Sex Female (\%) & 45.4 & 46.8 \\
			Race/Ethnicity (\%) & & \\
			\phantom{jj} White & 82.7 & 81.4 \\
			\phantom{jj} Black & 8.8 & 8.9 \\
			\phantom{jj} Hispanic & 4.2 & 4.4 \\
			\phantom{jj} Other & 4.2 & 5.3 \\
			Region (\%) & & \\
			\phantom{jj} Northeast & 20.3 & 20.2 \\
			\phantom{jj} Midwest & 13.6 & 13.2 \\
			\phantom{jj} West & 34.2 & 34.5 \\
			\phantom{jj} South & 31.9 & 32.2 \\
			Stage at Diagnosis (\%) & & \\
			\phantom{jj} I/II & 15.1 & 14.8 \\
			\phantom{jj} III & 34.0 & 33.1 \\
			\phantom{jj} IV & 50.8 & 52.0 \\ 
			\bottomrule 
	\end{tabular}}
\end{table*}
\end{center}

For stage classification, our input features are 94 variables derived from or linked to the Medicare claims data in the period 3 months before or after an individual receives their first lung cancer chemotherapy. Variables include patient demographic characteristics, visits and hospitalizations, chemotherapy drugs, surgeries and procedures, radiation, comorbidities, and lung cancer anatomic site and malignancy diagnosis codes. A full list of features for this data set is described in Brooks et al.\cite{Brooks19b} We examine patient survival to illustrate how our classification method can be used in practice with outcomes estimation. Survival outcomes are estimated based on the number of days from first chemotherapy to death, and we follow patients for a one year period.

Prior work classifying lung cancer stage for patients receiving chemotherapy focuses on a binary split of early (stages I-III) vs late (stage IV).\cite{Bergquist17,Brooks19b} Because it is unclear which single algorithm will perform best in the multiclass setting (stages I/II, stage III, and stage IV), we implement 7 algorithms that are multiclass versions of the most promising discrete binary prediction algorithms: main terms multinomial logistic regression\cite{Venables2002}; penalized regressions (lasso, ridge, an elastic net with overall $\lambda$ penalty selected via internal cross-validation, and a balanced elastic net penalty set at 0.5)\cite{Friedman2010}, generalized additive regression with cubic splines and a smoothing parameter set to 0.6\cite{Wood2011}; random forests with node size 250 and 500 trees\cite{Liaw2002}; and gradient boosting with a maximum tree depth of 3, learning rate set to 1, and 2 fitting rounds.\cite{Chen2016}

\subsection{Data analysis results}

In the weighted labeling approach, all algorithms provide at least nominal coverage (90\%) across all three stage classes. Under the naive method,  for stage I/II, the average coverage is about 31\%, stage III 60\%, and stage IV about 84\%. Table~\ref{tab:appbscov} contains the average coverage and 95\% confidence intervals across bootstrap samples for each algorithm and method.

\subsubsection{Classification performance}
The thresholds for the weighted labeling approach vary across algorithms and classes (Table~\ref{tab:appthlds}). All of the algorithms produce the lowest thresholds for stage I/II and the highest thresholds for stage IV. The stage IV thresholds are similar across all algorithms (ranging from 0.31--0.40), but the random forests generates the lowest thresholds for stage I/II (0.00 vs. 0.07--0.11) and stage III (0.07 vs 0.21--0.23). 

\begin{figure}[htbp]
	\centering
	\includegraphics[scale=.6]{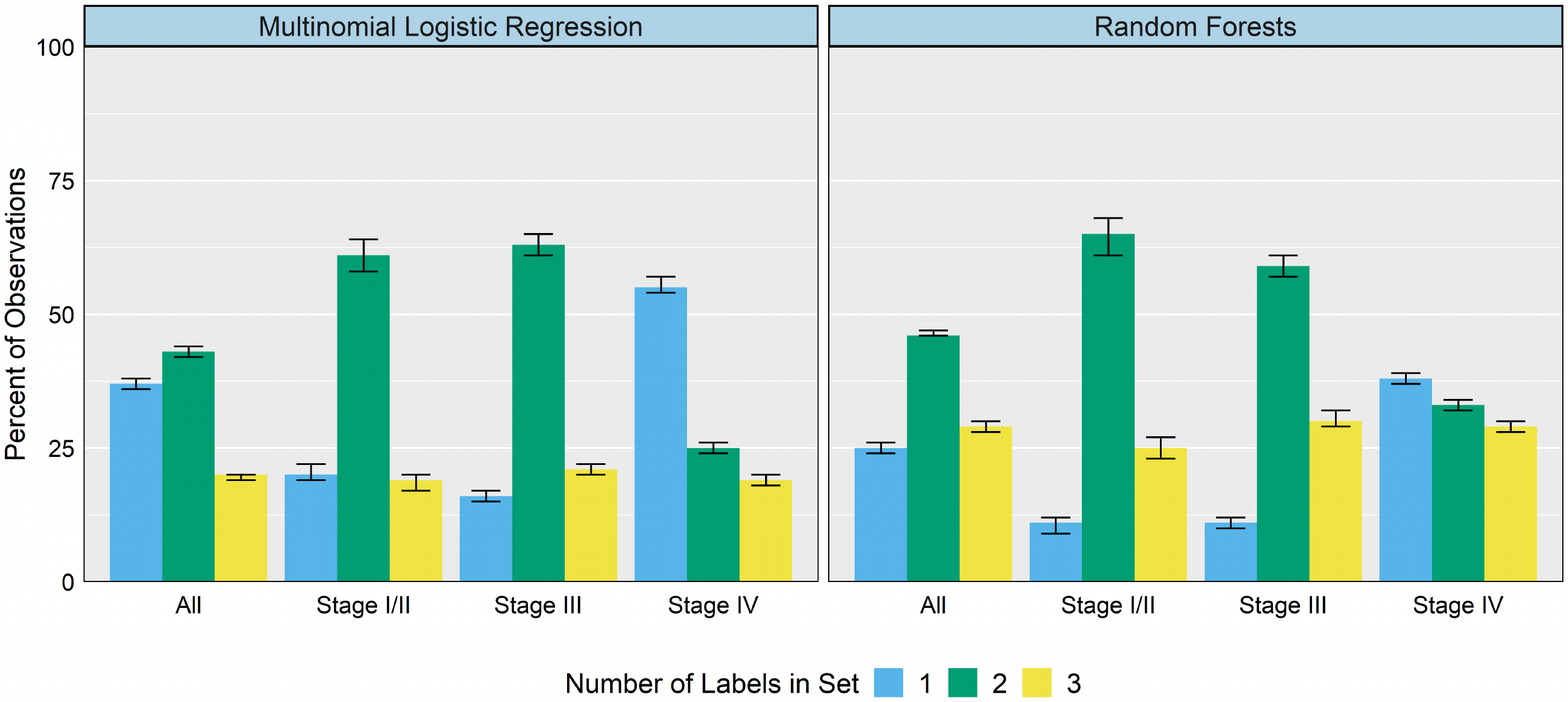}
%	\caption*{\footnotesize{Note: In the data analysis example, no null label sets were generated by any of the classification algorithms.}}
	\caption{Data analysis label ambiguity: Share of sample by number of assigned labels in label set.}
	\caption*{(\emph{Results for additional algorithms omitted due to similarity with multinomial logistic regression.})}
	\label{fig:dataamb}
\end{figure}

None of the algorithms produce a null label set (multinomial logistic regrssion and random forests presented in Figure~\ref{fig:dataamb}; results for other algorithms omitted due to similarity with multinomial logistic regression). Across all validation observations, random forests produces the fewest single label sets (about 24\% of validation observations) and the most triple label sets (30\%). In contrast, the multinomial logistic regression produces more single (35\%) and double label sets (44\%) across all classes. Examining ambiguity by true label class, we see the algorithms produce fewer single label sets for observations with a true class of stage I/II, reflecting the lower thresholds for this class. True stage IV is most likely to belong to a single label set, while true stages I/II and III are more likely to belong to a double label set, and all true classes have similar proportions of triple label sets. 

\begin{figure}[htbp]
	\centering
	\includegraphics[scale=.7]{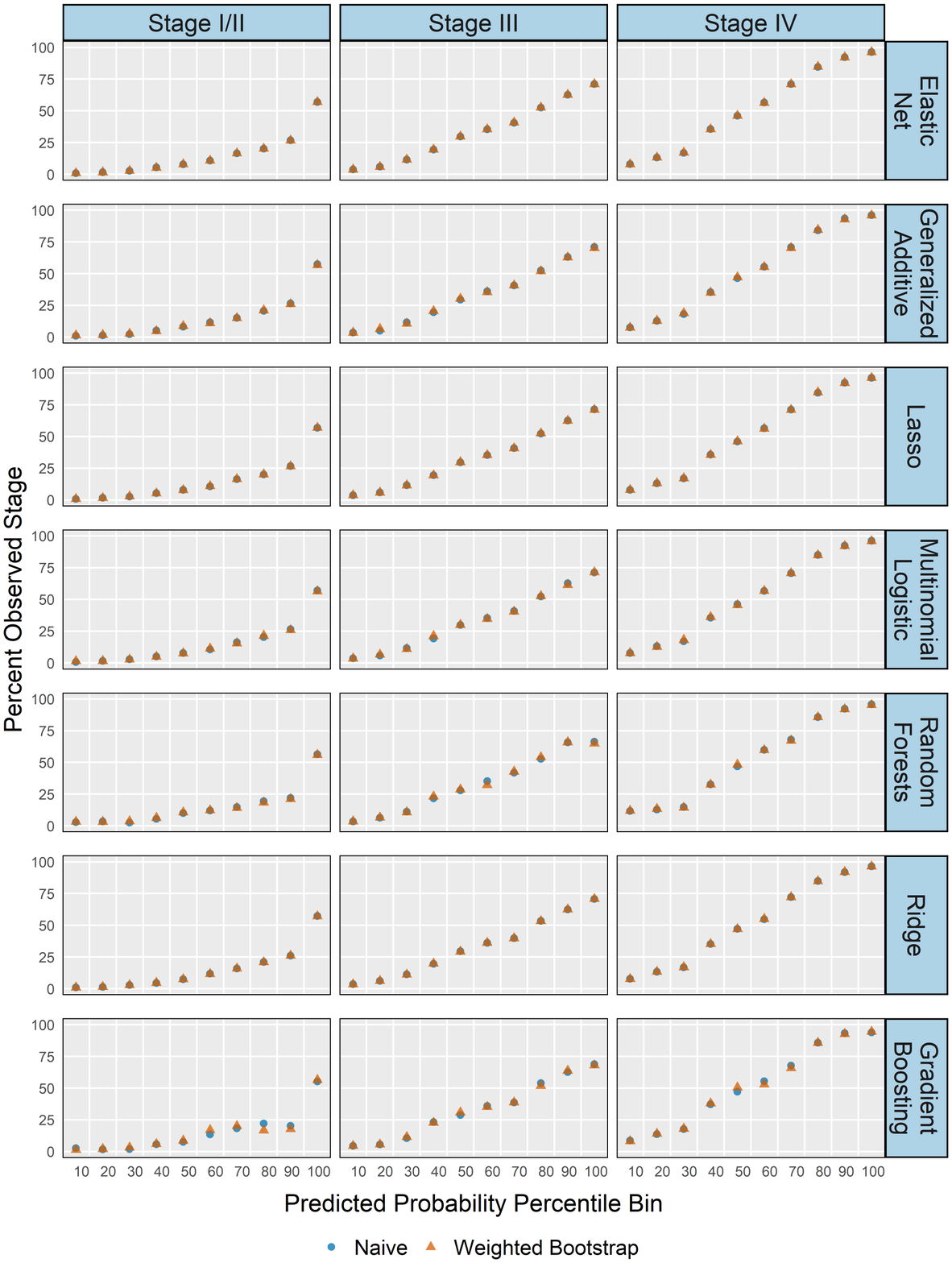}
	\caption{Data analysis: Observed stage by predicted probability.}
	\label{fig:appdatacal}
\end{figure}

To assess prediction calibration for each algorithm, we plot ordered predicted probabilities against the percent of observations belonging to a given stage (Figure~\ref{fig:appdatacal}). Recall the naive prediction methods use the entire development sample for algorithm fit, while the weighted bootstrap approach uses only half of the development sample for algorithm fitting, and the other half to set labeling thresholds. The average predicted probabilities across the distributions are nearly identical between the two methods for all stages and algorithms, with slight exceptions in gradient boosting and random forests. Overall, the algorithms are most poorly calibrated for stage I/II, show slightly better calibration for stage III, and perform best for stage IV.

\begin{figure}[htbp]
	\centering
	\includegraphics[scale=.6]{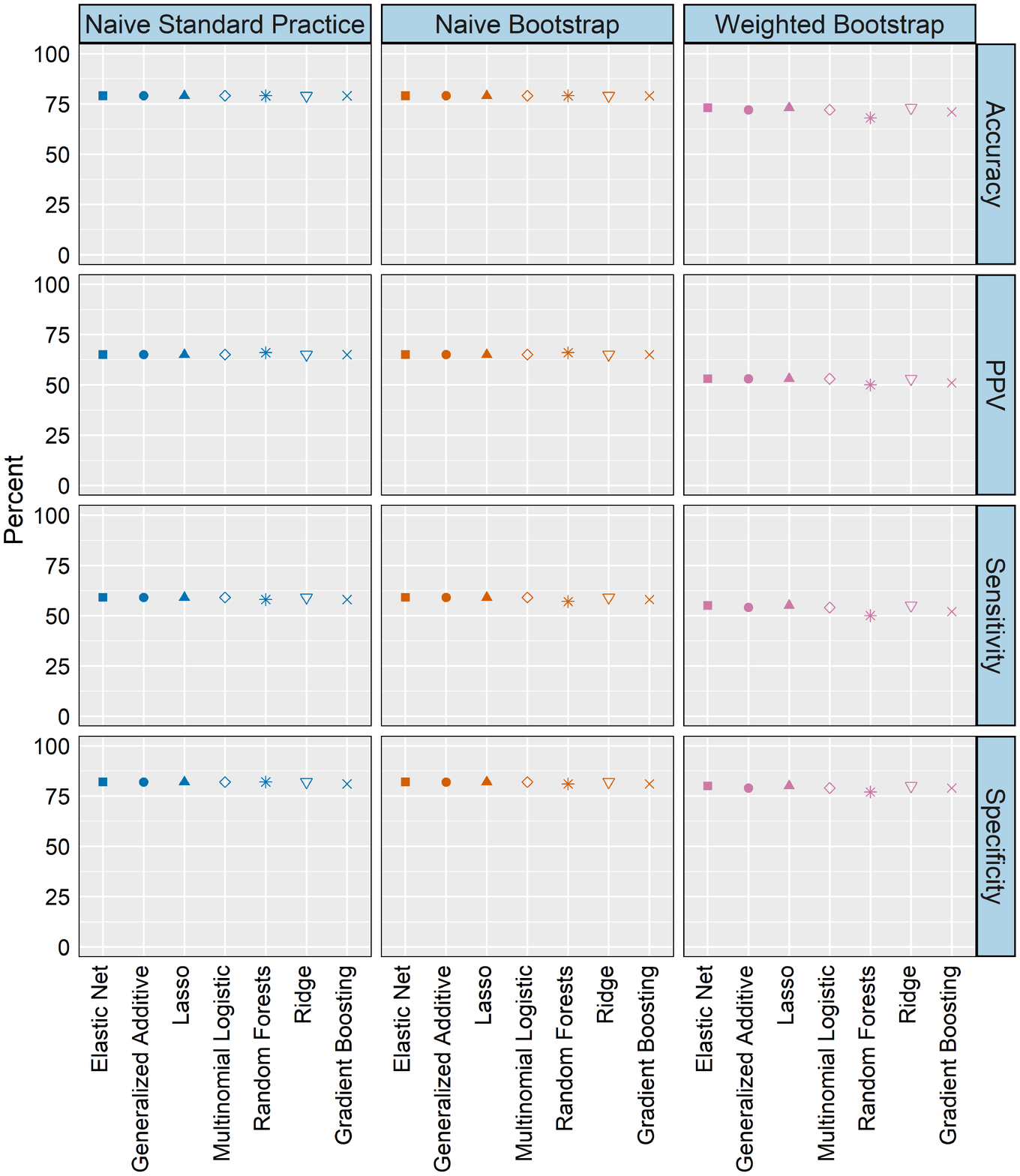}
	\caption{Data analysis classification discrimination: Average accuracy, sensitivity, specificity, and PPV.}
	\caption*{(\emph{For visual clarity, 95\% confidence intervals less than 0.05 are not displayed.})}
	\label{fig:datacmavg}
\end{figure}

Within the naive methods, the algorithms produce nearly identical classification performance results for several measures of discrimination (Figure~\ref{fig:datacmavg}). There are some small variations across algorithms within the weighted bootstrap method, but in general the classification performance of all algorithms---except the random forests---is within 1-2 percentage points. Tables~\ref{tab:appdataacc}-\ref{tab:appdatappv} present class-specific measures. Comparing the weighted bootstrap results to the naive methods, sensitivity is higher for stage I/II (by 12-19 percentage points), and sensitivity and PPV estimates are higher among stage IV. The weighted bootstrap yields slightly lower accuracy across all three stage groupings compared to the naive methods. Overall, in Figure~\ref{fig:appdatacal} and Figure~\ref{fig:datacmavg}, we see that measures of calibration and classification are similar, although in some cases there may be small differences, and performance by stage group is more variable.

\subsubsection{Survival analysis}

\begin{figure}[htbp]
	\centering
	\includegraphics[scale=.6]{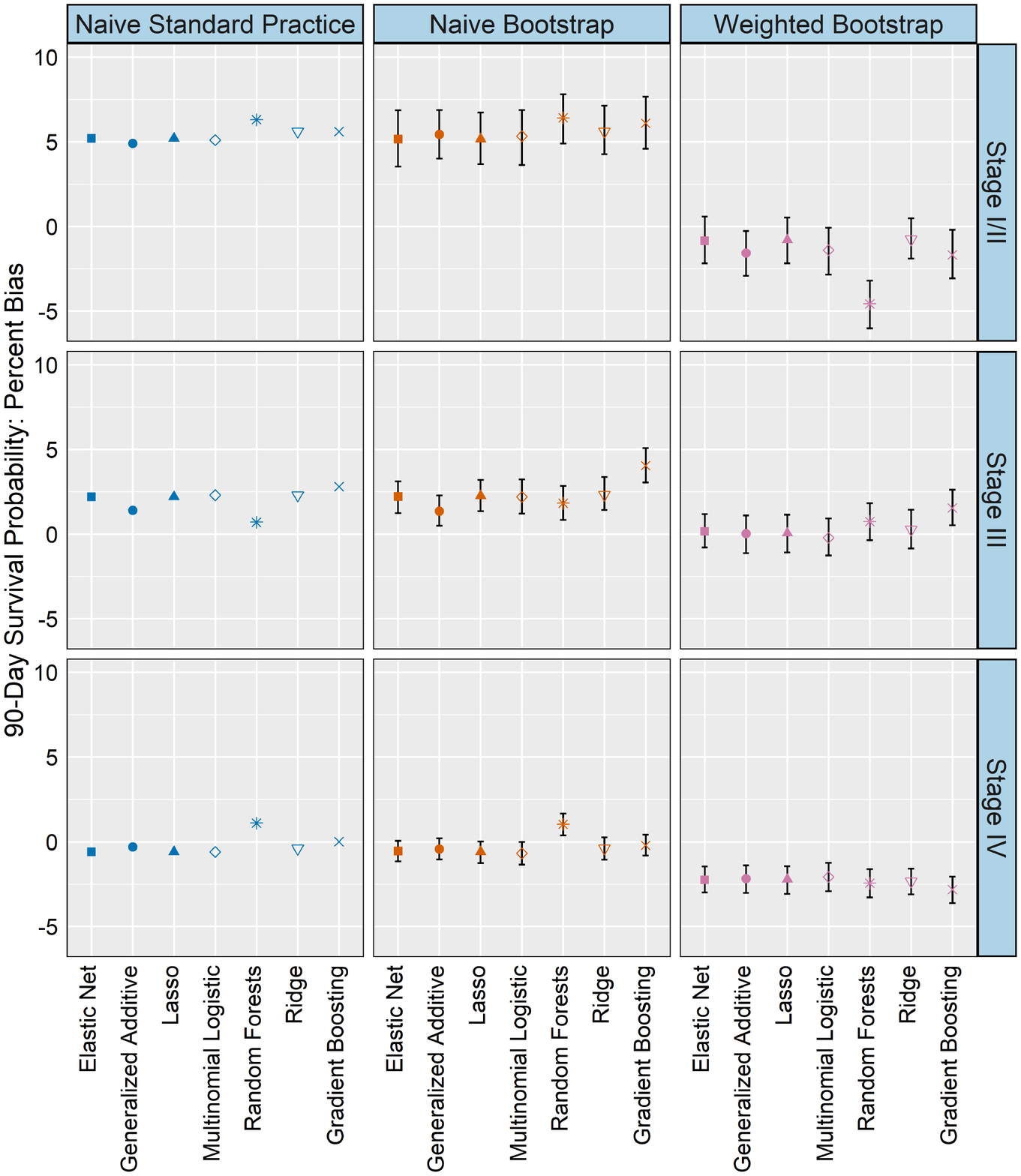}
	\caption{Data analysis: 90-day survival probability percent bias.}
	\label{fig:datad90bias}
\end{figure}

\begin{figure}[htbp]
	\centering
	\includegraphics[scale=.6]{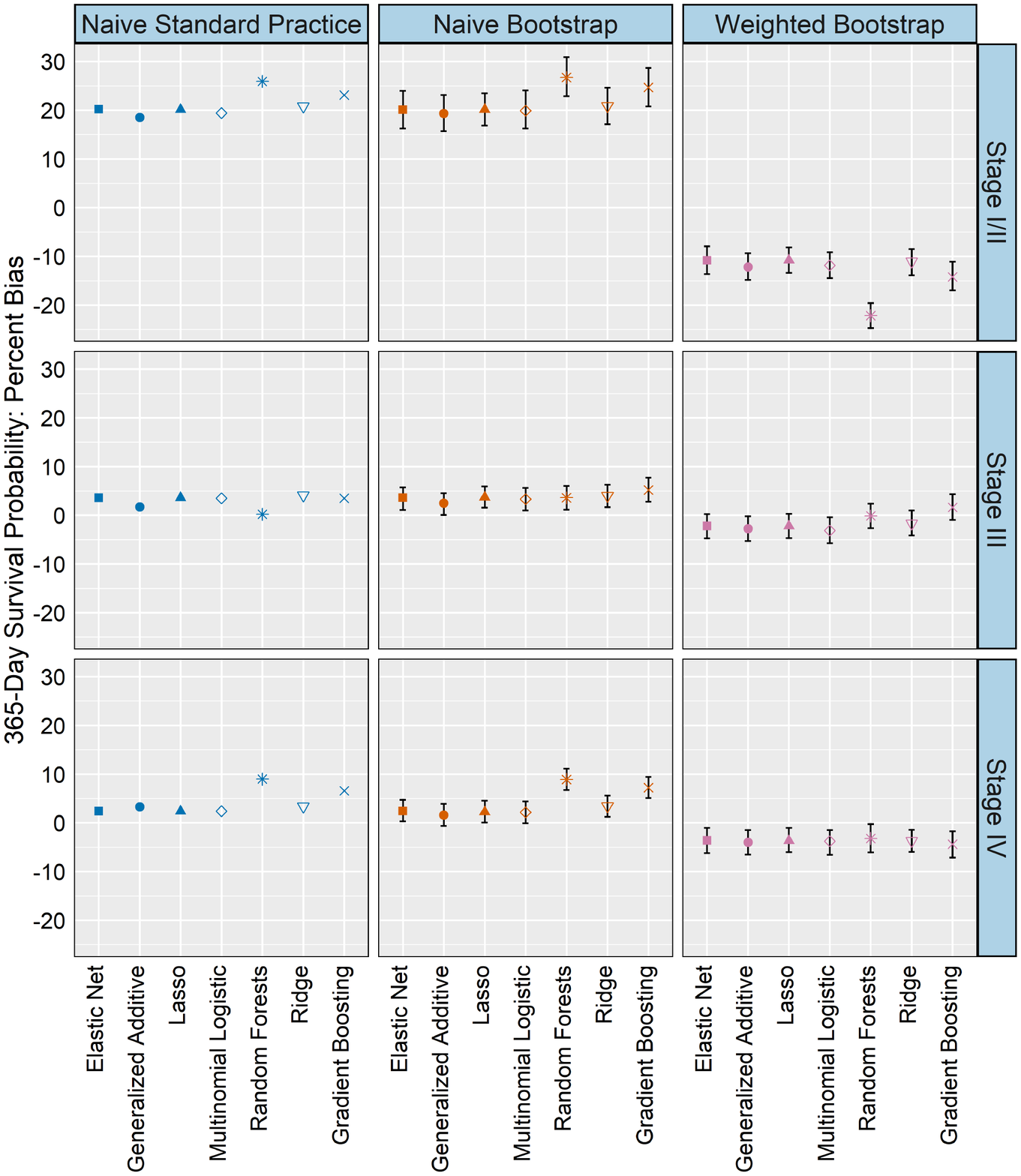}
	\caption{Data analysis: 365-day survival probability percent bias.}
	\label{fig:appdatad365bias}
\end{figure}

\begin{figure}[htbp]
	\centering
	\includegraphics[scale=.6]{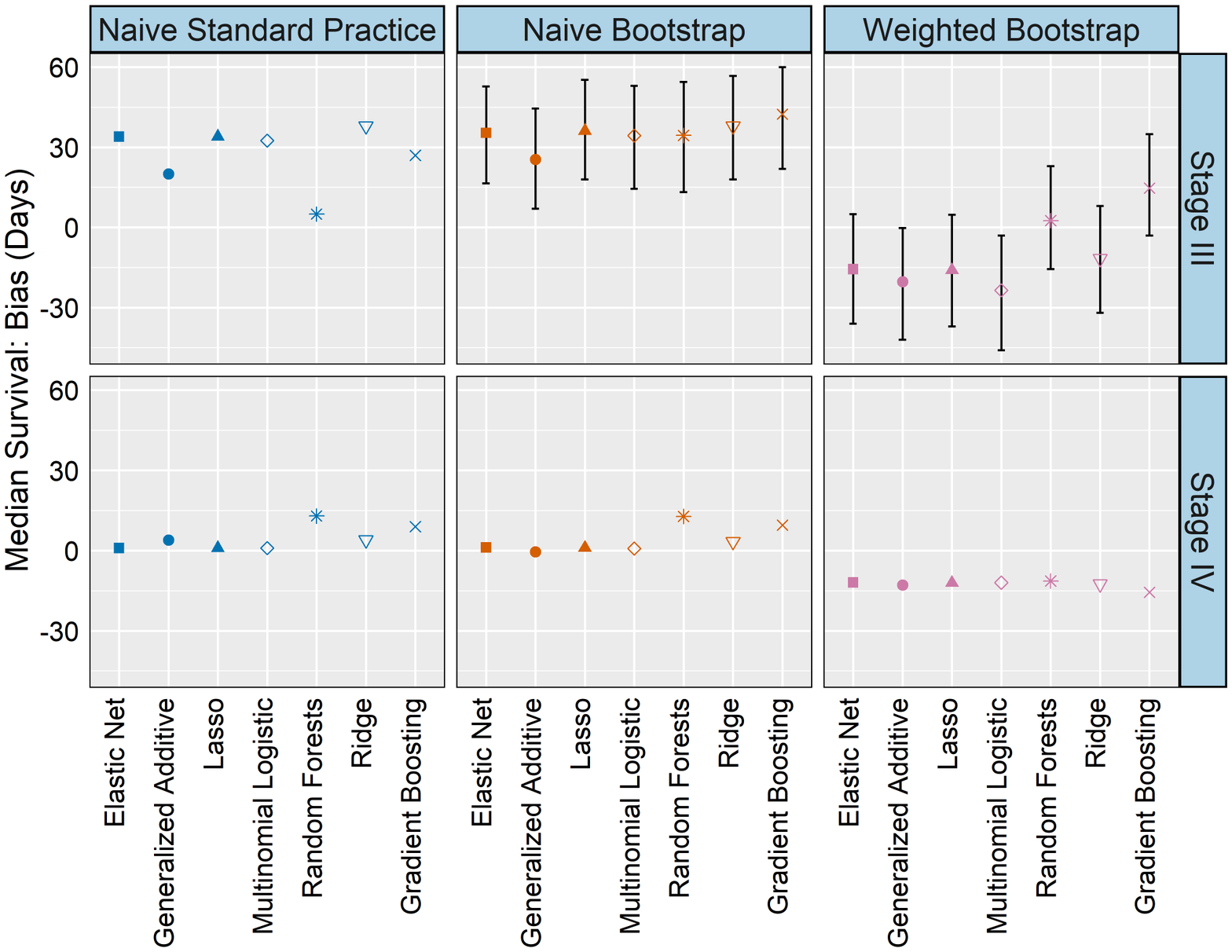}
	\caption{Data analysis: Median days survival bias.}
	\caption*{(\emph{For visual clarity, bootstrap-based 95\% confidence intervals less than 31 days are not displayed.})}
	\label{fig:appdatamedbias}
\end{figure}

For 90-day survival, the weighted bootstrap is close to 0\% bias across all stages, whereas the naive methods show greater variation (Figure~\ref{fig:datad90bias}). The weighted bootstrap is approximately 5 percentage points better than the naive methods in stage I/II, about 2 percentage points better in stage III, and only about 2 percentage points worse than the naive methods for stage IV. The naive and weighted labeling bootstrap approaches yield similarly sized bootstrap-based confidence intervals. For 365-day survival, the weighted labeling method continues to generate the smallest bias for stages I/II and III, and in some algorithms also produces the smallest bias in stage IV (Figure~\ref{fig:appdatad365bias}). The bias estimates are most different between the naive methods and the weighted label bootstrap for stage I/II, which is also the stage group that saw the greatest difference in classification results (Table~\ref{tab:appdataacc}-\ref{tab:appdatappv}). The observed median survival for stage I/II exceeds 1 year, so we do not estimate it here. For stage III, both naive methods tend to overestimate median survival compared to the true class, although they generate little bias for stage IV. The weighted labeling bootstrap method slightly underestimates median survival in both stage III and IV (Figure~\ref{fig:appdatamedbias}). 

\begin{figure}[htbp]
	\centering
	\includegraphics[scale=.6]{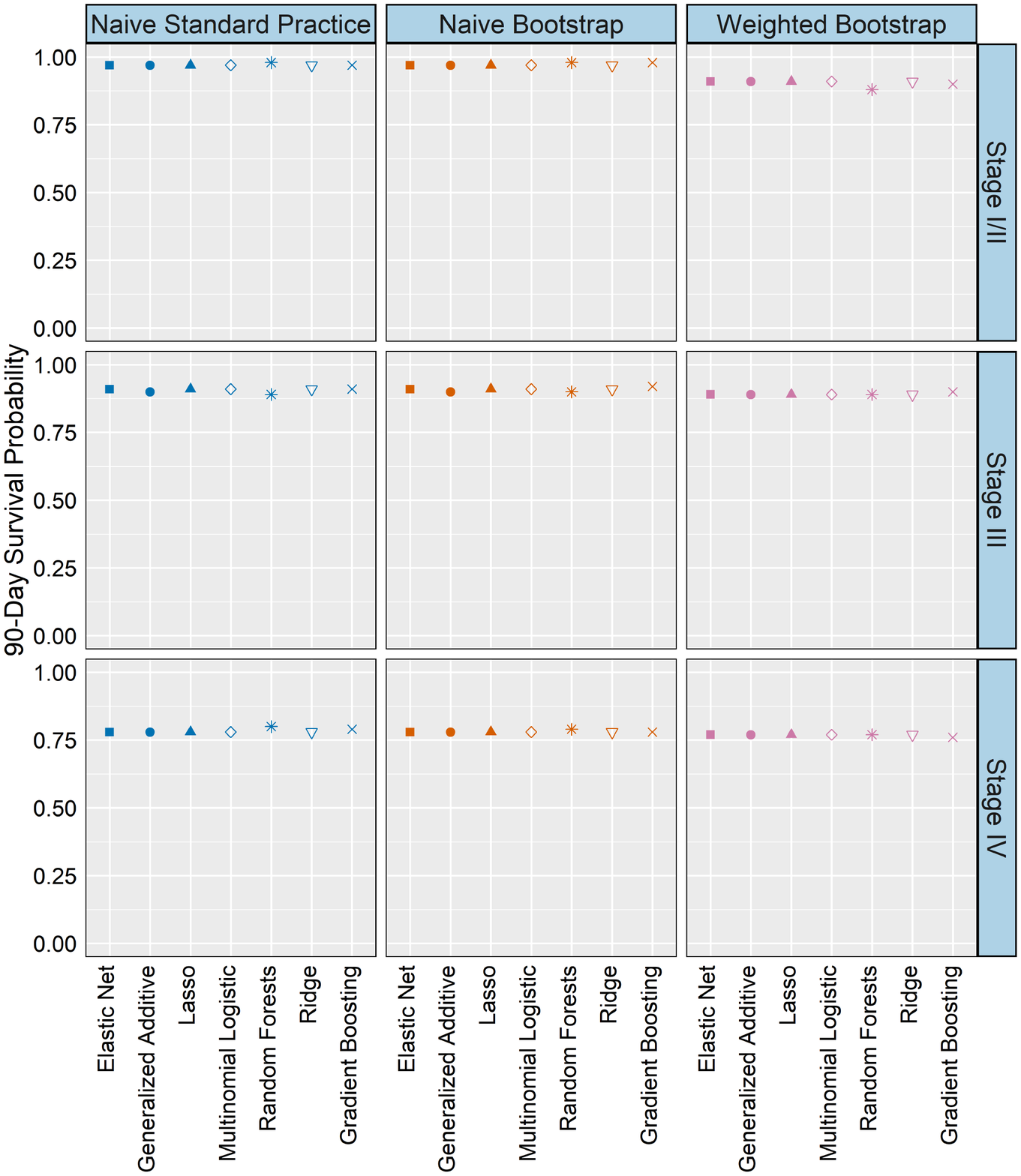}
	\caption{Data analysis: 90-day survival by predicted stage.}
	\caption*{(\emph{For visual clarity, 95\% confidence intervals less than 0.05 are not displayed.})}
	\label{fig:app90survpred}
\end{figure}

\begin{figure}[htbp]
	\centering
	\includegraphics[scale=.6]{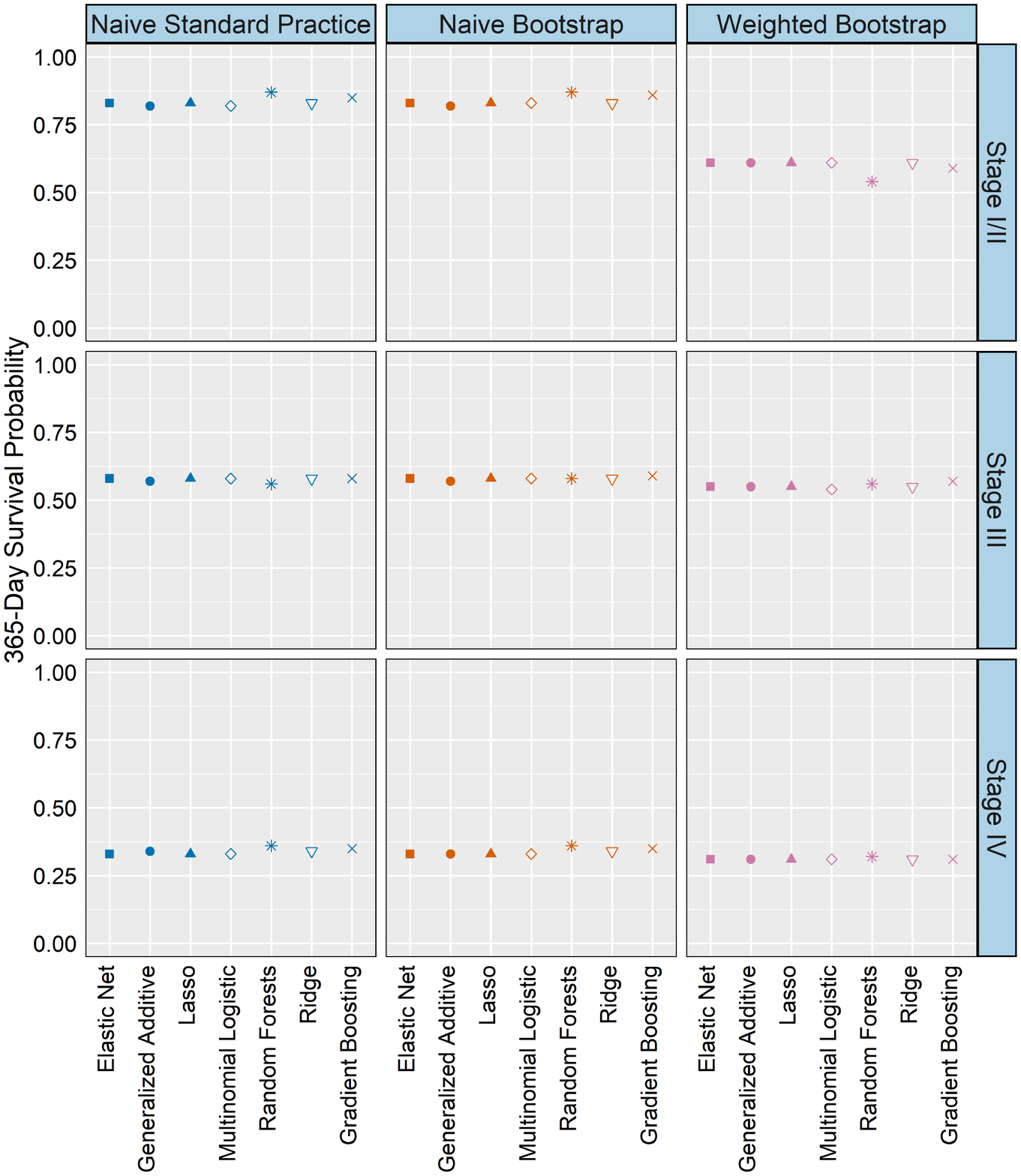}
	\caption{Data analysis: 365-day survival by predicted stage.}
	\caption*{(\emph{For visual clarity, 95\% confidence intervals less than 0.05 are not displayed.})}
	\label{fig:app365survpred}
\end{figure}

\begin{figure}[htbp]
	\centering
	\includegraphics[scale=.6]{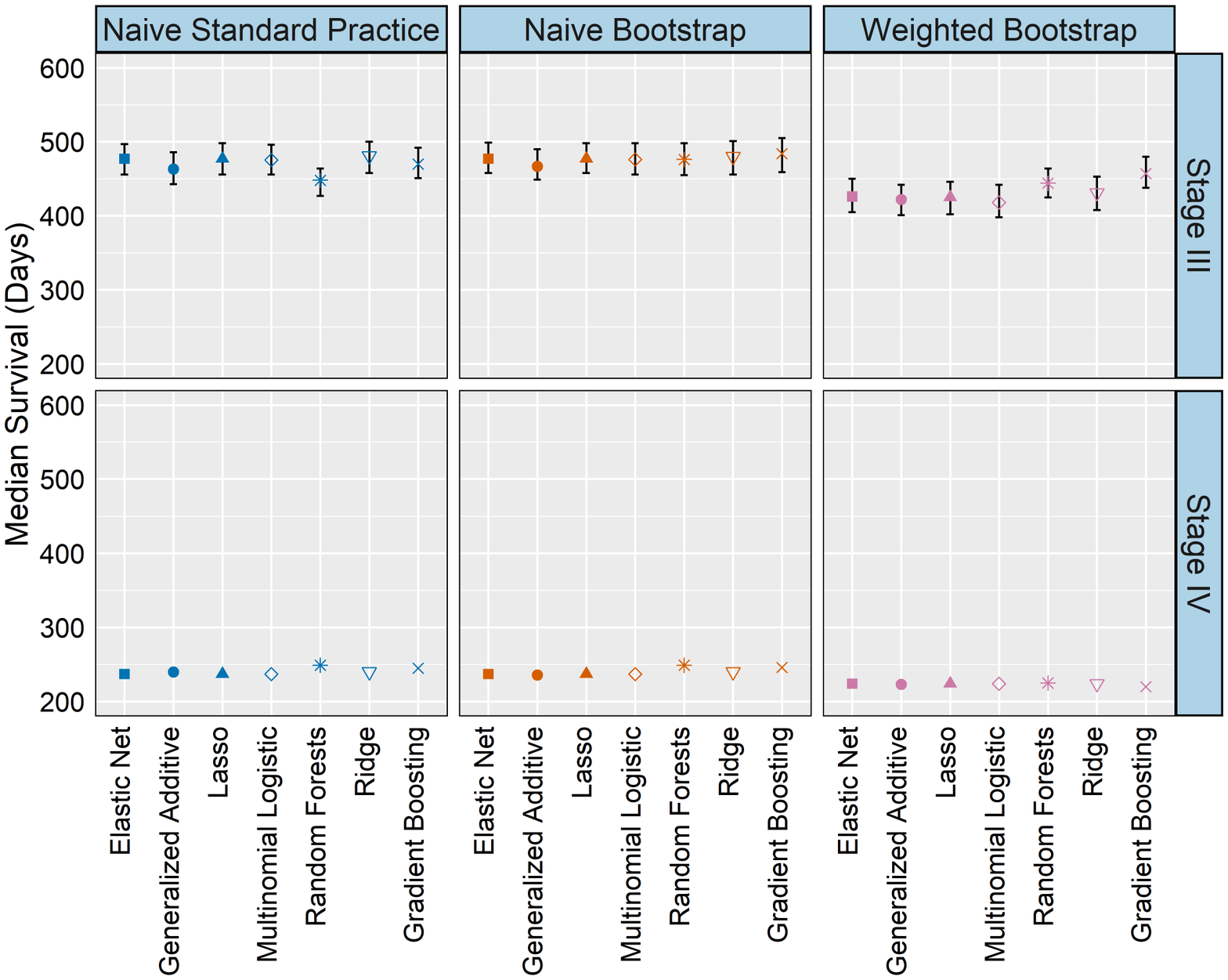}
	\caption{Data analysis: Median survival by predicted stage.}
	\caption*{(\emph{For visual clarity, 95\% confidence intervals less than 31 days are not displayed.})}
	\label{fig:appmedsurvpred}
\end{figure}

The naive standard practice and naive bootstrap methods generally produce similarly sized confidence intervals around the survival estimates, although there are some small variations in relative width across algorithms and class (see Figures~\ref{fig:app90survpred}-\ref{fig:appmedsurvpred}). Across all survival estimates the weighted labeling bootstrap method produces very similar results to the naive methods but does generate slightly narrower confidence intervals for the stage I/II 365-day survival estimates. Similar to the simulation study, we find the standard practice survival estimates do not necessarily yield smaller confidence intervals or less biased average survival probabilities than the naive bootstrap method. Instead, they tend to produce similar results that vary by predicted class and the algorithm used to estimate conditional probabilities. 

\section{Discussion}\label{sec:discussion}

In this article, we studied conveying uncertainty in applied classification settings, and proposed a procedure leveraging set-valued classification, split conformal inference, and resampling. Our proposed method uses bootstrap resampling from the sets of plausible labels generated in the classification step, and then performs outcomes estimation based on the selected labels. In our real-world data example, we developed fixed multiclass prediction algorithms for labeling lung cancer stage. The weighted labeling procedure yielded the smallest bias for survival estimates in stages I/II and stage III and was near 0\% bias for all three stages. 

Our development and validation samples had similar observed characteristics and were drawn from sequential time periods: development cohort patients were diagnosed with lung cancer from 2010--2011 and validation cohort patients in 2012--2013. As such, the naive boostrap approach may in practice be preferable to the weighted labeling bootstrap is such scenarios due to the simplicity of implementation without dramatic losses in performance. However, had our validation sample been drawn from a very different population---for example, patients diagnosed in 2019--2020---we might find much higher levels of uncertainty and lower levels of prediction accuracy due to changes in treatment patterns and patient characteristics. In this case, the weighted labeling approach may provide a more nuanced picture of the label uncertainty and impact on outcomes estimation by generating label sets for each observation.

Simulations show that our method outperforms the naive methods in terms of bias for some classes and scenarios, and for others yields similar survival estimates and confidence intervals. The simulation study also illustrates how label uncertainty can vary across classes and be translated into poor outcomes estimation performance; for example, class 3 yielded a higher number of ambiguous label sets and the greatest amount of bias across all simulation scenarios.  

The goal of this study was to propose an approach for characterizing uncertainty from a prediction exercise and demonstrate how this uncertainty can be incorporated in downstream survival analysis. As noted in the text, while we implemented a simple Kaplan-Meier analysis, there are more sophisticated time-to-event estimation approaches that could be deployed. Similarly, there are a range of other outcomes that may be of interest, including adjusted quality of care measures or cost-effectiveness estimates for oncology treatments. 

In summary, the weighted labeling with bootstrap method is one approach for incorporating uncertainty in applied classification and estimation problems. The proposed naive bootstrap procedure is also an improvement over simply ignoring prediction uncertainty. Depending on the data setting, the naive bootstrap may be implemented as an uncomplicated alternative to the weighted labeling method. As classification and risk prediction algorithms become more commonplace in medical and health services settings, we must think beyond prediction evaluation and implement tools to effectively communicate  uncertainty. 

%\backmatter

\section*{Acknowledgments}
This study used the linked SEER-Medicare database. The interpretation and reporting of these data are the sole responsibility of the authors. The authors acknowledge the efforts of the National Cancer Institute; the Office of Research, Development and Information, CMS; Information Management Services (IMS), Inc.; and the Surveillance, Epidemiology, and End Results (SEER) Program tumor registries in the creation of the SEER-Medicare database. The collection of cancer incidence data used in this study was supported by the California Department of Public Health as part of the statewide cancer reporting program mandated by California Health and Safety Code Section 103885; the National Cancer Institute's Surveillance, Epidemiology and End Results Program under contract HHSN261201000140C awarded to the Cancer Prevention Institute of California, contract HHSN261201000035C awarded to the University of Southern California, and contract HHSN261201000034C awarded to the Public Health Institute; and the Centers for Disease Control and Prevention's National Program of Cancer Registries, under agreement \# U58DP003862-01 awarded to the California Department of Public Health. The ideas and opinions expressed herein are those of the author(s) and endorsement by the State of California Department of Public Health, the National Cancer Institute, and the Centers for Disease Control and Prevention or their Contractors and Subcontractors is not intended nor should be inferred. The authors acknowledge the efforts of the National Cancer Institute; the Office of Research, Development and Information, CMS; Information Management Services (IMS), Inc.; and the Surveillance, Epidemiology, and End Results (SEER) Program tumor registries in the creation of the SEER-Medicare database.

\subsection*{Author contributions}
This is an author contribution text.

\subsection*{Financial disclosure}
None reported.

\subsection*{Conflict of interest}
The authors declare no potential conflict of interests.

\section*{Supporting information}
The following supporting information is available as part of the online article: Appendix A-C.

\clearpage
\newpage

\section*{Main Text Figures}

\newpage
\clearpage

\appendix

\section{Classification measures\label{sec:app0}}

\begin{center}
\begin{table}[htbp]
	\centering
	\caption{Classification performance measures.\label{tab:measures}}%
	{\begin{tabular}{  l  c  c  } 
			\toprule
			\multirow{2}{*}{\bf{Measure}} & \multicolumn{2}{ c  }{\bf{Definition}}  \\ 
			& \bf{Class-Specific} & \bf{Macro-Average} \\
			\midrule
			Accuracy &  $\sum_{i=1}^K \frac{tp_i + tn_i}{tp_i + tn_i + fp_i + fn_i}$ & $\sum_{i=1}^K \frac{tp_i + tn_i}{tp_i + tn_i + fp_i + fn_i} /K $  \\ 
			& & \\
			Sensitivity (Recall) & $\sum_{i=1}^K \frac{tp_i}{tp_i + fn_i}$ & $ \sum_{i =1}^K \frac{tp_i}{tp_i + fn_i}/K $\\ 
			& & \\
			Specificity & $\sum_{i=1}^K \frac{tn_i}{tn_i + fp_i}$ & $ \sum_{i =1}^K \frac{tn_i}{tn_i + fp_i}/K $  \\ 
			& & \\
			Positive Predictive Value (Precision) & $\sum_{i=1}^K \frac{tp_i}{tp_i + fp_i}$ & $ \sum_{i =1}^K \frac{tp_i}{tp_i + fp_i}/K $ \\
			\bottomrule
			\multicolumn{3}{ l } 
			{\begin{footnotesize}Notes: $tp_i$ denotes observation $i$ as true positive, $tn_i$ true negative,$fp_i$ false positive, $fn_i$ false negative,\end{footnotesize}} \\
			\multicolumn{2}{ l } 
			{\begin{footnotesize}$K$ is the number of class labels.\end{footnotesize}} \\
	\end{tabular}}
\end{table}
\end{center}

\newpage
\clearpage

\section{Simulation study}\label{app1}

Data are simulated using R 3.6.1 using a Mersenne-Twister random number generator with an input seed ``33.'' The outcome $Y_i$ is based on a multinomial logit where the multinomial probabilities are calculated as follows:

\begin{equation}\nonumber
	p_{ik} = 
	\begin{cases}
		\frac{\exp(X'_i b_1)}{1+\exp(X'_i b_1)+\exp(X'_ib_2)} & \text{for}\ k=1 \\
		\frac{\exp(X'_i b_2)}{1+\exp(X'_i b_1)+\exp(X'_ib_2)} & \text{for}\ k=2 \\
		\frac{1}{1+\exp(X'_i b_1)+\exp(X'_ib_2)} & \text{for}\ k=3, \\
	\end{cases}	
\end{equation}

\noindent where $X_i$ is a vector of predictor values for observation $i$ and $b_1$ and $b_2$ are vectors of coefficients corresponding to classes 1 and 2, respectively. We set the coefficients:

\begin{eqnarray}\nonumber
	X'_ib_1 = 1.8\times (-8.25 +0.2X1_i + 0.24(X7_i \times X10_i) - 0.3X3_i + 0.21\sqrt{X14_i} - 0.9X9_i + 0.9X11_i + 0.1\sin(X5_i) )
\end{eqnarray}

\begin{eqnarray}\nonumber
	X'_ib_2 \!=\! 1.8\times (-1.95 +0.04X1_i + 0.5(X7_i \times X10_i) - 0.03X3_i + 0.032\sqrt{X14_i} - 0.02X9_i + 0.003X11_i + 0.31\sin(X5_i) )
\end{eqnarray}

\begin{center}
	\begin{table}[htbp]%
		\centering
		\caption{Simulation covariates.\label{tab:appsim}}%
		\begin{tabular*}{500pt}{@{\extracolsep\fill}llcccc@{\extracolsep\fill}}
			\toprule
			& & & \multicolumn{3}{@{}c@{}}{\textbf{Used for Prediction}} \\\cmidrule{4-6}
			\textbf{Covariate} & \textbf{Distribution}  & \textbf{Used to Generate $Y_i$}  & \textbf{Scenario 1}  & \textbf{Scenario 2} & \textbf{Scenario 3} \\
			\midrule	
			X1 & N(75,5)  &  $\times$  & $\times$  & $\times$  &  \\	
			X2 & N(45000,10000)  &  &  & $\times$  &  $\times$ \\	
			X3 & N(23,4)  & $\times$  & $\times$   & $\times$ &  \\		
			X4 & N(70,5)  &   &   & $\times$ &  $\times$ \\	
			X5 & N(5,2)  & $\times$   & $\times$  & $\times$  &  $\times$ \\	
			X6 & N(0,1) \tnote{$\ddagger$} & &  $\times$  & $\times$  & $\times$  \\
			X7 & Bernoulli(0.5)  & $\times$ & $\times$  & &  $\times$ \\
			X8 & Bernoulli(0.25)  &   & $\times$ & & $\times$  \\
			X9 & Bernoulli(0.3)  & $\times$   & $\times$   &  & $\times$  \\
			X10 & Bernoulli(0.7)  &$\times$  & $\times$ & & $\times$  \\
			X11 & Bernoulli(0.6)  & $\times$   & $\times$ &  &  \\
			X12 & Bernoulli(0.7) &  &  & &  $\times$ \\
			X13 & Bernoulli(0.4)  &   & $\times$   & $\times$  & $\times$  \\		
			X14 & Pois(3)\tnote{$\dagger$} & $\times$  &   & &  $\times$ \\
			X15 & Pois(3)\tnote{$\dagger$} &  &   & & $\times$   \\		
			\bottomrule
		\end{tabular*}
		\begin{tablenotes}
			\item[$\dagger$] X14 and X15 count variables are correlated and based on a multivariate normal distribution: $\text{MVN}(\mu=(1,3),\Sigma)$, where $\Sigma= \big(\begin{smallmatrix}
				1 & .7 \\
				.7  & 1
			\end{smallmatrix}\big)$ 
			\item[$\ddagger$] X6 is the covariate used to generate survival times.
		\end{tablenotes}
	\end{table}
\end{center}

\newpage 
\clearpage

\subsection{Simulation results}

% bootstrap based coverege
\begin{center}
	\begin{table}[htbp]
		\centering
		\caption{Bootstrap-based coverage.\label{tab:appsimcov}}%
		\begin{tabular}{lccc}
			\toprule
			\textbf{Method} & \textbf{Class 1} & \textbf{Class 2} & \textbf{Class 3} \\ 
			\midrule
			\textit{Scenario 1} & & & \\
			Naive &  0.94 & 0.87 & 0.41 \\ 
			 & (0.92, 0.96) & (0.84, 0.90) & (0.32, 0.49) \\ 
			Weighted bootstrap & 0.90 & 0.90 & 0.90 \\ 
		   & (0.88, 0.93) & (0.88, 0.93) & (0.86, 0.96) \\ 
			\textit{Scenario 2} & & & \\
			Naive &  0.82 & 0.80 & 0.21 \\ 
			 & (0.78, 0.86) & (0.76, 0.84) & (0.14, 0.27) \\ 
			Weighted bootstrap & 0.90 & 0.90 & 0.91 \\ 
			 & (0.88, 0.93) & (0.88, 0.93) & (0.86, 0.96) \\ 
			\textit{Scenario 3} & & & \\
		 Naive& 0.54 & 0.62 & 0.00 \\ 
		& (0.49, 0.58) & (0.58, 0.67) & (0.00, 0.01) \\ 
		Weighted bootstrap & 0.90 & 0.90 & 0.91 \\
		 & (0.87, 0.93) & (0.88, 0.93) & (0.86, 0.96) \\ 
			\bottomrule
		\end{tabular}
	\end{table}
\end{center}

% weighted labeling thresholds
\begin{center}
	\begin{table}[htbp]
		\centering
		\caption{Weighted labeling thresholds.\label{tab:appsimthlds}}%
		\begin{tabular}{lrrr}
			\toprule
			\textbf{Scenario} & \textbf{Class 1} & \textbf{Class 2} & \textbf{Class 3} \\ 
			\midrule
1 & 0.659 & 0.393 & 0.112 \\ 
2& 0.290 & 0.321 & 0.102 \\ 
3 & 0.242 & 0.322 & 0.063 \\ 
			\bottomrule
		\end{tabular}
	\end{table}
\end{center}

% sim discrimination figure

% classification accuracy table
\begin{center}
	\begin{table}[htbp]
		\centering
		\caption{Classification accuracy. \label{tab:appsimacc}}%
		\begin{tabular}{lcccc}
			\toprule
			\textbf{Method} & \textbf{Average}  & \textbf{Class 1}  & \textbf{Class 2} & \textbf{Class 3}  \\ 
			\midrule
			\textit{Scenario 1} & & & & \\
			Naive standard practice & 0.89 & 0.95 & 0.84 & 0.89 \\ 
			& & & & \\
			Naive bootstrap  & 0.89 & 0.95 & 0.84 & 0.89 \\ 
			& (0.88, 0.91) & (0.94, 0.96) & (0.82, 0.86) & (0.87, 0.91) \\ 
			Weighted bootstrap & 0.86 & 0.94 & 0.80 & 0.84 \\ 
			& (0.85, 0.88) & (0.93, 0.96) & (0.77, 0.82) & (0.82, 0.86) \\ 
			\textit{Scenario 2} & & & & \\
			Naive standard practice & 0.82 & 0.86 & 0.73 & 0.87 \\ 
			& & & & \\
			Naive bootstrap & 0.82 & 0.86 & 0.73 & 0.87 \\ 
			 & (0.80, 0.84) & (0.83, 0.88) & (0.71, 0.76) & (0.85, 0.89) \\ 
			Weighted bootstrap & 0.77 & 0.84 & 0.66 & 0.80 \\ 
			 & (0.75, 0.79) & (0.82, 0.87) & (0.63, 0.69) & (0.78, 0.82) \\ 
			\textit{Scenario 3} & & & & \\
		Naive standard practice & 0.67 & 0.58 & 0.57 & 0.87 \\ 
		& & & & \\
		Naive bootstrap & 0.67 & 0.58 & 0.57 & 0.87 \\ 
	& (0.65, 0.69) & (0.55, 0.61) & (0.54, 0.60) & (0.85, 0.89) \\ 
		Weighted bootstrap & 0.57 & 0.56 & 0.53 & 0.63 \\ 
	 & (0.55, 0.59) & (0.53, 0.59) & (0.50, 0.56) & (0.60, 0.66) \\ 
		\hline
			\bottomrule
		\end{tabular}
	\end{table}
\end{center}

% classfiication sensitivity table
\begin{center}
	\begin{table}[htbp]
		\centering
		\caption{Classification sensitivty. \label{tab:appsimsens}}%
		\begin{tabular}{lcccc}
			\toprule
			\textbf{Method} & \textbf{Average}  & \textbf{Class 1}  & \textbf{Class 2} & \textbf{Class 3}  \\ 
			\midrule
			\textit{Scenario 1} & & & & \\
Naive standard practice& 0.74 & 0.94 & 0.87 & 0.41 \\ 
& & & & \\
Naive bootstrap & 0.74 & 0.94 & 0.87 & 0.41 \\ 
 & (0.71, 0.77) & (0.91, 0.96) & (0.84, 0.90) & (0.32, 0.49) \\ 
Weighted bootstrap & 0.75 & 0.91 & 0.75 & 0.61 \\ 
 & (0.72, 0.79) & (0.88, 0.93) & (0.71, 0.78) & (0.53, 0.69) \\ 
\textit{Scenario 2} & & & & \\
Naive standard practice & 0.61 & 0.82 & 0.80 & 0.21 \\ 
& & & & \\
Naive bootstrap & 0.61 & 0.82 & 0.80 & 0.21 \\ 
& (0.58, 0.64) & (0.78, 0.85) & (0.76, 0.83) & (0.14, 0.28) \\ 
Weighted bootstrap & 0.62 & 0.81 & 0.56 & 0.50 \\ 
 & (0.59, 0.66) & (0.77, 0.85) & (0.52, 0.61) & (0.41, 0.58) \\ 
\textit{Scenario 3} & & & & \\
Naive standard practice & 0.39 & 0.54 & 0.62 & 0.00 \\ 
& & & & \\
Naive bootstrap& 0.39 & 0.54 & 0.62 & 0.00 \\ 
& (0.37, 0.41) & (0.49, 0.58) & (0.58, 0.67) & (0.00, 0.01) \\ 
Weighted bootstrap& 0.36 & 0.35 & 0.37 & 0.36 \\ 
& (0.32, 0.39) & (0.30, 0.40) & (0.32, 0.41) & (0.27, 0.44) \\
			\bottomrule
		\end{tabular}
	\end{table}
\end{center}

% classification specificity table
\begin{center}
	\begin{table}[htbp]
		\centering
		\caption{Classification specificity. \label{tab:appsimspec}}%
		\begin{tabular}{lcccc}
			\toprule
			\textbf{Method} & \textbf{Average}  & \textbf{Class 1}  & \textbf{Class 2} & \textbf{Class 3}  \\ 
			\midrule
			\textit{Scenario 1} & & & & \\
		Naive standard practice & 0.91 & 0.96 & 0.81 & 0.96 \\ 
& & & & \\
		Naive bootstrap & 0.91 & 0.96 & 0.81 & 0.96 \\ 
		& (0.90, 0.92) & (0.94, 0.97) & (0.78, 0.84) & (0.95, 0.97) \\ 
		Weighted bootstrap & 0.90 & 0.97 & 0.85 & 0.88 \\ 
		 & (0.88, 0.91) & (0.95, 0.98) & (0.82, 0.88) & (0.86, 0.90) \\ 
		 \textit{Scenario 2} & & & & \\
		Naive standard practice & 0.84 & 0.88 & 0.67 & 0.97 \\ 
& & & & \\
		Naive bootstrap & 0.84 & 0.88 & 0.67 & 0.97 \\ 
		 & (0.83, 0.86) & (0.86, 0.91) & (0.63, 0.71) & (0.96, 0.98) \\ 
		Weighted bootstrap& 0.82 & 0.87 & 0.75 & 0.84 \\ 
		 & (0.80, 0.84) & (0.84, 0.89) & (0.71, 0.79) & (0.82, 0.87) \\ 
		 \textit{Scenario 3} & & & & \\
Naive standard practice& 0.71 & 0.61 & 0.52 & 1.00 \\ 
& & & & \\
Naive bootstrap& 0.71 & 0.61 & 0.52 & 1.00 \\ 
& (0.69, 0.72) & (0.57, 0.64) & (0.48, 0.56) & (1.00, 1.00) \\ 
Weighted bootstrap& 0.68 & 0.70 & 0.68 & 0.67 \\ 
& (0.67, 0.70) & (0.66, 0.73) & (0.64, 0.72) & (0.64, 0.70) \\ 
			\bottomrule
		\end{tabular}
	\end{table}
\end{center}

% classification ppv table
\begin{center}
	\begin{table}[htbp]
		\centering
		\caption{Classification positive predictive value. \label{tab:appsimppv}}%
		\begin{tabular}{lcccc}
			\toprule
			\textbf{Method} & \textbf{Average}  & \textbf{Class 1}  & \textbf{Class 2} & \textbf{Class 3}  \\ 
			\midrule
			\textit{Scenario 1} & & & & \\
Naive standard practice & 0.78 & 0.93 & 0.81 & 0.61 \\ 
& & & & \\
Naive bootstrap & 0.78 & 0.93 & 0.81 & 0.61 \\ 
 & (0.75, 0.82) & (0.91, 0.96) & (0.77, 0.84) & (0.51, 0.71) \\ 
Weighted bootstrap & 0.73 & 0.95 & 0.81 & 0.42 \\ 
 & (0.70, 0.76) & (0.93, 0.97) & (0.78, 0.85) & (0.35, 0.50) \\ 
\textit{Scenario 2} & & & & \\
Naive standard practice & 0.68 & 0.82 & 0.69 & 0.52 \\ 
& & & & \\
Naive bootstrap & 0.68 & 0.82 & 0.69 & 0.52 \\ 
 & (0.63, 0.73) & (0.78, 0.86) & (0.65, 0.73) & (0.38, 0.66) \\ 
Weighted bootstrap & 0.60 & 0.80 & 0.67 & 0.32 \\ 
 & (0.57, 0.63) & (0.76, 0.84) & (0.62, 0.72) & (0.26, 0.39) \\ 
\textit{Scenario 3} & & & & \\
Naive standard practice & 0.40 & 0.47 & 0.54 & 0.20 \\ 
& & & & \\
Naive bootstrap & 0.40 & 0.47 & 0.54 & 0.20 \\ 
& (0.35, 0.47) & (0.43, 0.52) & (0.50, 0.58) & (0.08, 0.37) \\ 
Weighted bootstrap & 0.36 & 0.43 & 0.50 & 0.14 \\ 
 & (0.33, 0.39) & (0.38, 0.49) & (0.45, 0.56) & (0.10, 0.17) \\ 
			\bottomrule
		\end{tabular}
	\end{table}
\end{center}

% true positive counts table
\begin{center}
	\begin{table}[htbp]
		\centering
		\caption{True positive counts.\label{tab:appsimtp}}%
		\begin{tabular}{lccc}
			\toprule
			\textbf{Method} & \textbf{Class 1} & \textbf{Class 2} & \textbf{Class 3} \\ 
			\midrule
			\textit{Scenario 1} & & & \\
			Naive standard practice & 374 & 414 & 52 \\ 
			 & & & \\
			Naive bootstrap & 374 & 414 & 52 \\ 
			 & (345, 404) & (383, 444) & (39, 67) \\
			Weighted bootstrap&  362 & 353 & 78 \\ 
			 &  (332, 391) & (324, 383) & (62, 95) \\ 
			\textit{Scenario 2} & & & \\
			Naive standard practice& 326 & 379 & 27 \\ 
			& & & \\
			Naive bootstrap & 326 & 379 & 27 \\ 
			& (297, 355) & (349, 409) & (17, 37) \\ 
		   Weighted bootstrap & 322 & 267 & 64 \\ 
			 &  (293, 351) & (240, 294) & (49, 79) \\ 
			 \textit{Scenario 3} & & & \\
			Naive standard practice &  214 & 295 & $<$1  \\ 
			& & & \\
			Naive bootstrap &  214 & 295 & $<$1  \\ 
			 & (189, 239) & (267, 323) & (0, 1) \\ 
			Weighted bootstrap &  140 & 173 & 46 \\ 
		 & (119, 161) & (150, 197) & (33, 59) \\
			\bottomrule
		\end{tabular}
	\end{table}
\end{center}

% true negative counts table
\begin{center}
	\begin{table}[htbp]
		\centering
		\caption{True negative counts.\label{tab:appsimtn}}%
		\begin{tabular}{lccc}
			\toprule
			\textbf{Method} & \textbf{Class 1} & \textbf{Class 2} & \textbf{Class 3} \\ 
			\midrule
			\textit{Scenario 1} & & & \\
			Naive standard practice & 575 & 427 & 838 \\ 
			& & &  \\
			Naive bootstrap& 575 & 427 & 838 \\ 
		 & (544, 605) & (397, 458) & (816, 861) \\ 
			Weighted bootstrap & 583 & 446 & 765 \\ 
		  &  (552, 613) & (415, 476) & (738, 790) \\ 
			\textit{Scenario 2} & & & \\
			Naive standard practice & 530 & 354 & 847 \\ 
			& & &  \\
			Naive bootstrap & 530 & 354 & 847 \\ 
			 & (500, 561) & (324, 383) & (825, 869) \\ 
			Weighted bootstrap& 522 & 394 & 736 \\ 
			&  (491, 552) & (365, 425) & (709, 763) \\ 
			\textit{Scenario 3} & & & \\
			Naive standard practice & 365 & 274 & 870 \\ 
			& & &  \\
			Naive bootstrap & 365 & 275 & 870 \\ 
			 &  (336, 395) & (247, 302) & (849, 890) \\ 
			Weighted bootstrap & 419 & 357 & 583 \\ 
			& (388, 449) & (328, 386) & (553, 613) \\ 
			\bottomrule
		\end{tabular}
	\end{table}
\end{center}

% false positive counts table
\begin{center}
	\begin{table}[htbp]
		\centering
		\caption{False positive counts.\label{tab:appsimfp}}%
		\begin{tabular}{lccc}
			\toprule
			\textbf{Method} & \textbf{Class 1} & \textbf{Class 2} & \textbf{Class 3} \\ 
			\midrule
			\textit{Scenario 1} & & & \\
		Naive standard practice& 27 & 99 & 34 \\ 
		& & &  \\
		Naive bootstrap & 27 & 99 & 34 \\ 
		& (17, 37) & (82, 118) & (23, 45) \\
		Weighted bootstrap& 19 & 81 & 107 \\ 
		 & (11, 28) & (64, 98) & (89, 127) \\ 
			\textit{Scenario 2} & & & \\
		 Naive standard practice & 71 & 173 & 25 \\ 
		& & &  \\
	Naive bootstrap & 71 & 173 & 25 \\ 
		 &(56, 87) & (150, 196) & (16, 35) \\ 
		Weighted bootstrap &  80 & 132 & 136 \\ 
		& (64, 97) & (111, 153) & (115, 157) \\ 
			\textit{Scenario 3} & & & \\
	Naive standard practice & 236 & 253 & 2 \\ 
		& & &  \\
		Naive bootstrap & 236 & 253 & 2 \\ 
		&  (211, 262) & (226, 280) & (0, 3) \\ 
		Weighted bootstrap & 182 & 170 & 289 \\
	&  (159, 206) & (148, 194) & (261, 317) \\ 
			\bottomrule
		\end{tabular}
	\end{table}
\end{center}

% false negative counts table
\begin{center}
	\begin{table}[htbp]
		\centering
		\caption{False negative counts.\label{tab:appsimfn}}%
		\begin{tabular}{lccc}
			\toprule
			\textbf{Method} & \textbf{Class 1} & \textbf{Class 2} & \textbf{Class 3} \\ 
			\midrule
			\textit{Scenario 1} & & & \\
			Naive standard practice & 24 & 60 & 76 \\ 
			& & &  \\
			Naive bootstrap & 24 & 60 & 76 \\ 
			 & (15, 34) & (46, 75) & (60, 92) \\ 
			Weighted bootstrap & 37 & 120 & 50 \\ 
			 & (26, 49) & (100, 140) & (37, 64) \\ 
			\textit{Scenario 2} & & & \\
			Naive standard practice &  72 & 95 & 101 \\ 
			& & &  \\
			Naive bootstrap &  72 & 95 & 101 \\ 
			 & (57, 89) & (77, 113) & (83, 120) \\ 
			Weighted bootstrap & 76 & 207 & 64 \\ 
			 &  (61, 93) & (182, 232) & (50, 80) \\ 
			\textit{Scenario 3} & & & \\
		 Naive standard practice& 185 & 178 & 128 \\ 
		& & &  \\
		Naive bootstrap & 185 & 178 & 128 \\ 
		 & (161, 209) & (155, 202) & (108, 149) \\ 
		Weighted bootstrap &  259 & 300 & 83 \\ 
		 & (232, 286) & (272, 328) & (66, 100) \\ 
			\bottomrule
		\end{tabular}
	\end{table}
\end{center}

% 90 day bias 

% 365-day bias

%%%%%%%% DATA ANALYSIS APPENDIX SECTION %%%%%%%%%
\newpage 
\clearpage

\section{Data analysis results}\label{app2}

%  Bootstrap based coverage
\begin{center}
\begin{table}[htbp]
	\centering
	\caption{Bootstrap-based coverage.\label{tab:appbscov}}%
	\begin{tabular}{lccc}
		\toprule
		\textbf{Algorithm} & \textbf{Stage I/II} & \textbf{Stage III} & \textbf{Stage IV} \\ 
		\midrule
		\textit{Weighted Labeling} & & & \\
Elastic Net & 0.89 & 0.90 & 0.90 \\ 
 & (0.88, 0.90) & (0.89, 0.90) & (0.89, 0.90) \\ 
Generalized Additive Regression & 0.90 & 0.90 & 0.89 \\ 
& (0.89, 0.92) & (0.89, 0.91) & (0.89, 0.90) \\ 
Lasso &  0.89 & 0.90 & 0.90 \\ 
 &  (0.88, 0.90) & (0.89, 0.91) & (0.89, 0.90) \\ 
Multinomial Logistic &0.90 & 0.90 & 0.89 \\ 
 & (0.89, 0.91) & (0.90, 0.91) & (0.89, 0.90) \\ 
Random Forests & 0.94 & 0.91 & 0.90 \\ 
 & (0.93, 0.95) & (0.91, 0.92) & (0.90, 0.91) \\ 
Ridge &  0.90 & 0.90 & 0.90 \\ 
 & (0.89, 0.91) & (0.89, 0.90) & (0.89, 0.91) \\ 
Gradient Boosting &  0.91 & 0.91 & 0.90 \\ 
 & (0.90, 0.92) & (0.90, 0.92) & (0.90, 0.91) \\ 
		\textit{Naive Labeling} & & & \\
Elastic Net &  0.32 & 0.61 & 0.84 \\ 
& (0.30, 0.34) & (0.60, 0.63) & (0.83, 0.85) \\ 
Generalized Additive Regression &  0.32 & 0.63 & 0.83 \\ 
& (0.30, 0.34) & (0.61, 0.64) & (0.83, 0.84) \\ 
Lasso & 0.32 & 0.61 & 0.84 \\ 
& (0.30, 0.34) & (0.60, 0.63) & (0.83, 0.85) \\ 
Multinomial Logistic & 0.32 & 0.61 & 0.84 \\ 
 & (0.30, 0.34) & (0.60, 0.62) & (0.83, 0.85) \\  
Random Forests&  0.30 & 0.58 & 0.87 \\ 
 &  (0.30, 0.34) & (0.59, 0.62) & (0.84, 0.86) \\ 
Ridge &  0.31 & 0.61 & 0.85 \\ 
& (0.29, 0.33) & (0.59, 0.62) & (0.84, 0.86) \\ 
Gradient Boosting & 0.30 & 0.58 & 0.86 \\ 
 & (0.28, 0.32) & (0.60, 0.63) & (0.84, 0.85) \\ 
		\bottomrule
	\end{tabular}
\end{table}
\end{center}

% weighted labeling thresholds
\begin{center}
	\begin{table}[htbp]
		\centering
		\caption{Weighted labeling thresholds.\label{tab:appthlds}}%
		\begin{tabular}{lrrr}
			\toprule
			\textbf{Algorithm} & \textbf{Stage I/II} & \textbf{Stage III} & \textbf{Stage IV} \\ 
			\midrule
		Elastic Net & 0.09 & 0.23 & 0.34 \\ 
		Generalized Additive Regression & 0.07 & 0.21 & 0.31 \\ 
		Lasso & 0.09 & 0.23 & 0.35 \\ 
		Multinomial Logistic & 0.07 & 0.21 & 0.34 \\ 
		Random Forests& 0.00 & 0.07 & 0.40 \\ 
		Ridge & 0.09 & 0.23 & 0.36 \\ 
		Gradient Boosting& 0.11 & 0.21 & 0.40 \\ 
			\bottomrule
		\end{tabular}
	\end{table}
\end{center}

% classification accuracy
\begin{center}
	\begin{table}[ht]
		\centering
		\caption{Classification accuracy. \label{tab:appdataacc}}%
		\begin{tabular}{lcccc}
			\toprule
			\textbf{Algorithm} & \textbf{Average}  & \textbf{Stage I/II}  & \textbf{Stage III} & \textbf{Stage IV}  \\ 
			\midrule
			\textit{Naive standard practice} & & & & \\
			Elastic Net & 0.79 & 0.87 & 0.73 & 0.78 \\ 
			Generalized Additive Regression & 0.79 & 0.87 & 0.73 & 0.78 \\ 
			Lasso & 0.79 & 0.87 & 0.73 & 0.78 \\ 
			Multinomial Logistic & 0.79 & 0.87 & 0.73 & 0.78 \\ 
			Random Forests & 0.79 & 0.87 & 0.73 & 0.77 \\ 
			Ridge & 0.79 & 0.87 & 0.73 & 0.78 \\ 
			Gradient Boosting & 0.79 & 0.87 & 0.74 & 0.77 \\ 
			\textit{Naive bootstrap} & & & & \\
		Elastic Net & 0.79 & 0.87 & 0.73 & 0.78 \\ 
		 & (0.79, 0.80) & (0.86, 0.87) & (0.72, 0.74) & (0.77, 0.78) \\ 
		Generalized Additive Regression & 0.79 & 0.87 & 0.73 & 0.78 \\ 
		 & (0.79, 0.80) & (0.86, 0.87) & (0.72, 0.74) & (0.77, 0.78) \\ 
		Lasso & 0.79 & 0.87 & 0.73 & 0.78 \\ 
		 & (0.79, 0.80) & (0.86, 0.87) & (0.72, 0.74) & (0.77, 0.78) \\ 
		Multinomial Logistic & 0.79 & 0.87 & 0.73 & 0.77 \\ 
		 & (0.79, 0.80) & (0.86, 0.87) & (0.72, 0.73) & (0.77, 0.78) \\ 
		Random Forests & 0.79 & 0.87 & 0.74 & 0.77 \\ 
		 & (0.79, 0.80) & (0.86, 0.87) & (0.73, 0.74) & (0.76, 0.78) \\ 
		Ridge & 0.79 & 0.87 & 0.73 & 0.78 \\ 
		 & (0.79, 0.80) & (0.86, 0.88) & (0.72, 0.74) & (0.77, 0.78) \\ 
		Gradient Boosting & 0.79 & 0.87 & 0.73 & 0.76 \\ 
		 & (0.78, 0.79) & (0.86, 0.87) & (0.73, 0.74) & (0.75, 0.77) \\ 
			 \textit{Weighted bootstrap} & & & & \\
			Elastic Net & 0.73 & 0.75 & 0.67 & 0.75 \\ 
			 & (0.72, 0.73) & (0.75, 0.76) & (0.67, 0.68) & (0.75, 0.76) \\ 
			Generalized Additive Regression & 0.72 & 0.74 & 0.67 & 0.75 \\ 
			 & (0.71, 0.72) & (0.73, 0.75) & (0.66, 0.67) & (0.74, 0.75) \\ 
			Lasso& 0.73 & 0.75 & 0.67 & 0.75 \\ 
			 & (0.72, 0.73) & (0.75, 0.76) & (0.66, 0.68) & (0.75, 0.76) \\ 
			Multinomial Logistic & 0.72 & 0.74 & 0.67 & 0.74 \\ 
			 & (0.71, 0.72) & (0.74, 0.75) & (0.66, 0.67) & (0.74, 0.75) \\ 
			Random Forests & 0.68 & 0.67 & 0.66 & 0.71 \\ 
			 & (0.67, 0.68) & (0.66, 0.68) & (0.65, 0.67) & (0.70, 0.71) \\ 
			Ridge & 0.73 & 0.75 & 0.67 & 0.76 \\ 
			 & (0.72, 0.73) & (0.74, 0.76) & (0.66, 0.68) & (0.75, 0.76) \\ 
			Gradient Boosting & 0.71 & 0.73 & 0.66 & 0.75 \\ 
			 & (0.71, 0.72) & (0.72, 0.74) & (0.66, 0.67) & (0.74, 0.75) \\ 
			\bottomrule
		\end{tabular}
	\end{table}
\end{center}

% classification sensitivity
\begin{center}
	\begin{table}[ht]
		\centering
		\caption{Classification sensitivity. \label{tab:appdatasens}}%
		\begin{tabular}{lcccc}
			\toprule
			\textbf{Algorithm} & \textbf{Average}  & \textbf{Stage I/II}  & \textbf{Stage III} & \textbf{Stage IV}  \\ 
			\midrule
			\textit{Naive standard practice} & & & & \\
			Elastic Net & 0.59 & 0.32 & 0.61 & 0.84 \\ 
			Generalized Additive Regression & 0.59 & 0.32 & 0.63 & 0.83 \\ 
			Lasso & 0.59 & 0.32 & 0.61 & 0.84 \\ 
			Multinomial Logistic & 0.59 & 0.32 & 0.61 & 0.84 \\ 
			Random Forests & 0.58 & 0.31 & 0.58 & 0.87 \\ 
			Ridge & 0.59 & 0.31 & 0.61 & 0.85 \\ 
			Gradient Boosting & 0.58 & 0.30 & 0.58 & 0.86 \\ 
			\textit{Naive bootstrap} & & & & \\
		Elastic Net & 0.59 & 0.32 & 0.61 & 0.84 \\ 
		 & (0.58, 0.60) & (0.30, 0.34) & (0.60, 0.63) & (0.83, 0.85) \\ 
		Generalized Additive Regression & 0.59 & 0.33 & 0.63 & 0.83 \\ 
		 & (0.59, 0.60) & (0.31, 0.35) & (0.62, 0.65) & (0.82, 0.83) \\ 
		Lasso & 0.59 & 0.32 & 0.61 & 0.84 \\ 
		 & (0.58, 0.60) & (0.30, 0.34) & (0.60, 0.63) & (0.83, 0.85) \\ 
		Multinomial Logistic & 0.59 & 0.32 & 0.62 & 0.83 \\ 
		 & (0.58, 0.60) & (0.30, 0.34) & (0.60, 0.63) & (0.83, 0.84) \\ 
		Random Forests & 0.57 & 0.28 & 0.56 & 0.88 \\ 
		 & (0.57, 0.58) & (0.26, 0.30) & (0.54, 0.57) & (0.88, 0.89) \\ 
		Ridge & 0.59 & 0.31 & 0.61 & 0.85 \\ 
		 & (0.58, 0.60) & (0.29, 0.33) & (0.59, 0.62) & (0.84, 0.86) \\ 
		Gradient Boosting & 0.58 & 0.3 & 0.55 & 0.88 \\ 
		 & (0.57, 0.58) & (0.28, 0.32) & (0.53, 0.56) & (0.87, 0.88) \\ 
			\textit{Weighted bootstrap} & & & & \\
		Elastic Net & 0.55 & 0.49 & 0.46 & 0.70 \\ 
		 & (0.54, 0.56) & (0.47, 0.51) & (0.44, 0.47) & (0.69, 0.71) \\ 
		Generalized Additive Regression & 0.54 & 0.49 & 0.45 & 0.68 \\ 
		 & (0.53, 0.55) & (0.47, 0.51) & (0.44, 0.47) & (0.67, 0.69) \\ 
		Lasso & 0.55 & 0.49 & 0.46 & 0.70 \\ 
		 & (0.54, 0.56) & (0.47, 0.51) & (0.44, 0.47) & (0.69, 0.71) \\ 
		Multinomial Logistic & 0.54 & 0.49 & 0.45 & 0.68 \\ 
		 & (0.53, 0.55) & (0.47, 0.51) & (0.44, 0.47) & (0.67, 0.69) \\ 
		Random Forests & 0.50 & 0.46 & 0.44 & 0.59 \\ 
		 & (0.49, 0.50) & (0.44, 0.48) & (0.42, 0.45) & (0.58, 0.60) \\ 
		Ridge & 0.55 & 0.48 & 0.45 & 0.70 \\ 
		 & (0.54, 0.56) & (0.46, 0.50) & (0.44, 0.47) & (0.69, 0.71) \\ 
		Gradient Boosting & 0.52 & 0.42 & 0.44 & 0.69 \\ 
		 & (0.51, 0.53) & (0.40, 0.44) & (0.43, 0.46) & (0.68, 0.70) \\ 
			\bottomrule
		\end{tabular}
	\end{table}
\end{center}

% classification specificity
\begin{center}
	\begin{table}[ht]
		\centering
		\caption{Classification specificity. \label{tab:appdataspec}}%
		\begin{tabular}{lcccc}
			\toprule
			\textbf{Algorithm} & \textbf{Average}  & \textbf{Stage I/II}  & \textbf{Stage III} & \textbf{Stage IV}  \\ 
			\midrule
			\textit{Naive standard practice} & & & & \\
			Elastic Net & 0.82 & 0.96 & 0.79 & 0.71 \\ 
			Generalized Additive Regression & 0.82 & 0.96 & 0.78 & 0.72 \\ 
			Lasso & 0.82 & 0.96 & 0.79 & 0.71 \\ 
			Multinomial Logistic & 0.82 & 0.96 & 0.78 & 0.71 \\ 
			Random Forests & 0.82 & 0.97 & 0.81 & 0.67 \\ 
			Ridge & 0.82 & 0.97 & 0.79 & 0.70 \\ 
			Gradient Boosting & 0.81 & 0.97 & 0.81 & 0.66 \\ 
			\textit{Naive bootstrap} & & & & \\
		Elastic Net & 0.82 & 0.96 & 0.79 & 0.71 \\ 
		 & (0.82, 0.82) & (0.96, 0.97) & (0.78, 0.79) & (0.7, 0.72) \\ 
		Generalized Additive Regression & 0.82 & 0.96 & 0.78 & 0.72 \\ 
		 & (0.82, 0.83) & (0.96, 0.97) & (0.77, 0.78) & (0.71, 0.73) \\ 
		Lasso & 0.82 & 0.96 & 0.79 & 0.71 \\ 
		 & (0.82, 0.82) & (0.96, 0.97) & (0.78, 0.79) & (0.7, 0.72) \\ 
		Multinomial Logistic & 0.82 & 0.96 & 0.78 & 0.71 \\ 
		 & (0.82, 0.82) & (0.96, 0.97) & (0.77, 0.79) & (0.7, 0.72) \\ 
		Random Forests & 0.81 & 0.97 & 0.82 & 0.64 \\ 
		 & (0.81, 0.82) & (0.97, 0.97) & (0.82, 0.83) & (0.63, 0.65) \\ 
		Ridge & 0.82 & 0.97 & 0.79 & 0.70 \\ 
		 & (0.82, 0.82) & (0.96, 0.97) & (0.78, 0.8) & (0.69, 0.71) \\ 
		Gradient Boosting & 0.81 & 0.97 & 0.83 & 0.64 \\ 
		 & (0.81, 0.81) & (0.96, 0.97) & (0.82, 0.83) & (0.62, 0.65) \\ 
			\textit{Weighted bootstrap} & & & & \\
		Elastic Net & 0.80 & 0.80 & 0.78 & 0.81 \\ 
		 & (0.79, 0.80) & (0.79, 0.81) & (0.77, 0.79) & (0.8, 0.82) \\ 
		Generalized Additive Regression & 0.79 & 0.79 & 0.77 & 0.82 \\ 
		& (0.79, 0.80) & (0.78, 0.79) & (0.76, 0.78) & (0.81, 0.82) \\ 
		Lasso & 0.80 & 0.80 & 0.78 & 0.81 \\ 
		 & (0.79, 0.80) & (0.79, 0.80) & (0.77, 0.79) & (0.80, 0.82) \\ 
		Multinomial Logistic & 0.79 & 0.79 & 0.77 & 0.81 \\ 
		 & (0.79, 0.80) & (0.78, 0.79) & (0.76, 0.78) & (0.81, 0.82) \\ 
		Random Forests & 0.77 & 0.71 & 0.77 & 0.83 \\ 
		 & (0.77, 0.78) & (0.70, 0.72) & (0.76, 0.78) & (0.82, 0.84) \\ 
		Ridge & 0.80 & 0.80 & 0.78 & 0.81 \\ 
		 & (0.79, 0.80) & (0.79, 0.80) & (0.77, 0.79) & (0.80, 0.82) \\ 
		Gradient Boosting & 0.79 & 0.78 & 0.77 & 0.80 \\ 
		 & (0.78, 0.79) & (0.78, 0.79) & (0.76, 0.78) & (0.80, 0.81) \\
			\bottomrule
		\end{tabular}
	\end{table}
\end{center}

% classification ppv
\begin{center}
	\begin{table}[ht]
		\centering
		\caption{Classification positive predictive value. \label{tab:appdatappv}}%
		\begin{tabular}{lcccc}
			\toprule
			\textbf{Algorithm} & \textbf{Average}  & \textbf{Stage I/II}  & \textbf{Stage III} & \textbf{Stage IV}  \\ 
			\midrule
			\textit{Naive standard practice} & & & & \\
			Elastic Net & 0.65 & 0.61 & 0.59 & 0.76 \\ 
			Generalized Additive Regression & 0.65 & 0.61 & 0.59 & 0.76 \\ 
			Lasso& 0.65 & 0.61 & 0.59 & 0.76 \\ 
		   Multinomial Logistic & 0.65 & 0.61 & 0.59 & 0.76 \\ 
			Random Forests & 0.66 & 0.63 & 0.60 & 0.74 \\ 
			Ridge & 0.65 & 0.62 & 0.59 & 0.75 \\ 
			Gradient Boosting & 0.65 & 0.62 & 0.61 & 0.73 \\ 
			\textit{Naive bootstrap} & & & & \\
		Elastic Net & 0.65 & 0.61 & 0.59 & 0.76 \\ 
		 & (0.64, 0.66) & (0.59, 0.64) & (0.57, 0.60) & (0.75, 0.76) \\ 
		Generalized Additive Regression & 0.65 & 0.60 & 0.58 & 0.76 \\ 
		 & (0.64, 0.66) & (0.57, 0.63) & (0.57, 0.60) & (0.76, 0.77) \\ 
		Lasso & 0.65 & 0.61 & 0.59 & 0.76 \\ 
		 & (0.64, 0.66) & (0.58, 0.64) & (0.57, 0.60) & (0.75, 0.77) \\ 
		Multinomial Logistic & 0.65 & 0.60 & 0.59 & 0.76 \\ 
		 & (0.64, 0.66) & (0.57, 0.63) & (0.57, 0.60) & (0.75, 0.77) \\ 
		Random Forests & 0.66 & 0.63 & 0.61 & 0.73 \\ 
		 & (0.64, 0.67) & (0.60, 0.66) & (0.60, 0.63) & (0.72, 0.74) \\ 
		Ridge & 0.65 & 0.62 & 0.59 & 0.75 \\ 
		 & (0.64, 0.67) & (0.59, 0.65) & (0.58, 0.60) & (0.74, 0.76) \\ 
		Gradient Boosting & 0.65 & 0.62 & 0.61 & 0.72 \\ 
		 & (0.64, 0.66) & (0.59, 0.65) & (0.60, 0.63) & (0.71, 0.73) \\
			\textit{Weighted bootstrap} & & & & \\
		Elastic Net & 0.53 & 0.30 & 0.51 & 0.80 \\ 
		 & (0.53, 0.54) & (0.28, 0.31) & (0.49, 0.52) & (0.79, 0.81) \\ 
		Generalized Additive Regression & 0.53 & 0.28 & 0.50& 0.08 \\ 
		 & (0.52, 0.53) & (0.27, 0.30) & (0.48, 0.51) & (0.79, 0.81) \\ 
		Lasso & 0.53 & 0.30 & 0.51 & 0.80 \\ 
		 & (0.53, 0.54) & (0.28, 0.31) & (0.49, 0.52) & (0.79, 0.81) \\ 
		Multinomial Logistic & 0.53 & 0.29 & 0.50 & 0.80 \\ 
		 & (0.52, 0.53) & (0.27, 0.30) & (0.48, 0.51) & (0.79, 0.81) \\ 
		Random Forests & 0.50 & 0.22 & 0.49 & 0.79 \\ 
		 & (0.49, 0.51) & (0.20, 0.23) & (0.47, 0.50) & (0.78, 0.80) \\ 
		Ridge & 0.53 & 0.29 & 0.50 & 0.80 \\ 
		 & (0.53, 0.54) & (0.28, 0.31) & (0.49, 0.52) & (0.79, 0.81) \\ 
		Gradient Boosting & 0.51 & 0.25 & 0.49 & 0.79 \\ 
		 & (0.51, 0.52) & (0.24, 0.27) & (0.48, 0.51) & (0.78, 0.80) \\ 
			\bottomrule
		\end{tabular}
	\end{table}
\end{center}

% calibration

%% Bias

%% Survival estimates and intervals

%\nocite{*}% Show all bib entries - both cited and uncited; comment this line to view only cited bib entries;

\newpage
\clearpage

\bibliography{ml4h_bib}%

\end{document}